\documentclass[useAMS,usenatbib]{mn2e}

\usepackage{verbatim,graphicx,epsfig,dcolumn,color}
\definecolor{red}{rgb}{0.75,0.0,0.0}
\definecolor{green}{rgb}{0.0,0.75,0.0}
\definecolor{blue}{rgb}{0.0,0.0,0.75}
\newcolumntype{.}{D{.}{.}{4}}
\newcolumntype{,}{D{.}{.}{2}}
\newcolumntype{;}{D{.}{.}{1}}
\newcommand{\nodata}{$\cdot\cdot\cdot$}
\newcommand{\lesssim}{{\lower-1.2pt\vbox{\hbox{\rlap{$<$}\lower5pt\vbox{\hbox{$\sim$}}}}}}
\newcommand{\gtrsim}{{\lower-1.2pt\vbox{\hbox{\rlap{$>$}\lower5pt\vbox{\hbox{$\sim$}}}}}}

\title[Globular clusters cleaned by white dwarfs]{Ejection of globular cluster interstellar media through ionization by white dwarfs}
\author[I. McDonald et al.]{I.~McDonald$^{1}$\thanks{E-mail: mcdonald@jb.man.ac.uk}, A.~A.~Zijlstra$^{1}$ \\
$^{1}$Jodrell Bank Centre for Astrophysics, Alan Turing Building, Manchester, M13 9PL, UK\\
}

\begin{document}

\date{Accepted 9999 December 32. Received 9999 December 32; in original form 9999 December 32}

\pagerange{\pageref{firstpage}--\pageref{lastpage}} \pubyear{9999}

\maketitle

\label{firstpage}

\begin{abstract}

UV radiation from white dwarfs can efficiently clear Galactic globular clusters (GCs) of their intra-cluster medium (ICM). This solves the problem of the missing ICM in clusters, which is otherwise expected to build up to easily observable quantities. To show this, we recreate the ionizing flux in 47 Tuc, following randomly generated stars through their AGB, post-AGB and white dwarf evolution. Each white dwarf can ionize all the material injected into the cluster by stellar winds for $\sim$3 Myr of its evolution: $\sim$40 such white dwarfs exist at any point. Every GC's ICM should be ionized. The neutral cloud in M15 should be caused by a temporary overdensity. A pressure-supported ICM will expand over the cluster's tidal radius, where it will be truncated, allowing Jeans escape. The modelled Jeans mass-loss rate approximates the total stellar mass-loss rate, allowing efficient clearing of ICM. Any cluster's ICM mass should equal the mass injected by its stars over the sound-travel time between the cluster core and tidal radius. We predict $\sim$11.3 M$_\odot$ of ICM within 47 Tuc, cleared over $\sim$4 Myr, compared to a dynamical timescale of 4.3 Myr. We present a new mass hierarchy, discussing the transition between globular clusters dwarf galaxies.
\end{abstract}



\section{Introduction}
\label{IntroSect}

The interstellar media of globular clusters are best known for their absence. The simple assumption was that matter lost by stars would build up in the centre of clusters, as it does in larger bodies (e.g.\ \citealt{VF77}). The absence of recent star formation alone requires that some mechanism must remove the intra-cluster medium (ICM) from globular clusters, but the real mechanisms and timescales are only now starting to become clear.

Secure detections of neutral ICM are limited to a single cluster: an H {\sc i} cloud of $\approx$0.3 M$_\odot$ with $9 \pm 2 \times 10^{-4}$ M$_\odot$ of dust detected in M15 \citep{ESvL+03,vLSEM06,BWvL+06}. Extensive searches for neutral interstellar gas and dust have otherwise provided only upper limits, with typical respective limits of $\sim$0.1--1 M$_\odot$ and $10^{-4}$ M$_\odot$ (e.g.\ \citealt{SWFW90,MMN+08,BMvL+08,BBW+09,vLSP+09}). An ionized medium of $\approx$0.1 M$_\odot$ has been detected in 47 Tucanae via dispersion measures towards millisecond pulsars \citep{FKL+01}. This remains the only detection of ionized ICM, but shows that the ICM can exist both as an ionized and neutral medium.

The lack of observed ICM necessitates a relatively short timescale for its dispersal. The mass-loss rates of individual giant stars in globular clusters are well-studied, and dust-producing stars can lose up to $\sim$10$^{-6}$ M$_\odot$ yr$^{-1}$ (e.g.\ \citealt{BMvL+09,MvLD+09,MBvLZ11,MvLS+11,SMM+10}). Stars not producing dust also appear to sustain substantial mass-loss rates (e.g.\ \citealt{Cohen76,DHA84,CBR+04,MvL07,DSS09}). Indeed, substiantial mass loss is required to take a main-sequence turn-off star of $\approx$0.8--0.9 M$_\odot$ and convert it into a white dwarf of $\approx$0.53 M$_\odot$ (\citealt{RFI+97,MKZ+04,KSDR+09,Kalirai13}; McDonald et al., in prep). One star is expected to go through this process of stellar death roughly once per $88\,000$ years for a 10$^6$ M$_\odot$ cluster (cf.\ \citealt{MBvL+11}, scaling linearly with the mass of the cluster), requiring a removal of neutral material from some clusters on timescales of $\lesssim$1 Myr.

\subsection{Energy sources for clearing clusters}

A variety of methods have been proposed for removing ICM from globular clusters. Material should be removed by ram pressure stripping as the cluster plunges through the Galactic Plane, but this only occurs every $\sim$100 Myr \citep{Roberts60,Roberts86,TW75}. Ram pressure stripping of gas as the cluster passes through the Galactic Halo has been a leading suggestion for removing ICM, but Halo gas appears to be of too low a density to effectively strip material from within all but the least massive clusters \citep{FG76,PRS11}.

It is now thought that an internal method must be clearing ICM. Existing theories can be grouped into episodic phenomena, which provide stochastic clearing of ICM, and continuous phenomena. The least-frequent mechanism may be stellar collisions \citep{UCR08}, which would be too infrequent to account for ICM clearing on timescales much shorter than every $\sim$10$^6$ years. \citet{SD78} proposed novae as a clearing mechanism and, while dwarf novae seem to be rare (e.g.\ \citealt{SWL+11}), they may be a plausible mechanism for small clusters \citep{MB11}. \citet{CW77} suggested M-dwarf flaring could clear clusters, but they probably under-estimated the effects of mass segregation on lower-mass stars (cf.\ \citealt{PRP+10}).

Continuous internal mechanisms for clearing ICM rely primarily on radiative or kinetic energy input from stellar winds. Sources for these include pulsars \citep{Spergel91,FKL+01}, which we address here; hot horizontal branch stars \citep{VF77}, which do not occur in every cluster; main-sequence-star winds \citep{Smith99,NSFRR13}; and fast winds from less-evolved red giants \citep{SDS04}. A primary concern for less-evolved stellar winds is that the momentum transfer from the wind must be efficiently thermalized for this mechanism to work ($\gtrsim$15 per cent efficiency; \citealt{NSFRR13}).

\subsection{Physical mechanisms of escape}
\label{PhysEscSect}

Ignoring ram pressure stripping by Halo gas, ICM can thermally escape from the cluster via one of three methods. {\it Jeans escape} occurs when the gas is sufficiently tenuous that it becomes collisionless: outward-moving particles with thermal velocities greater than the escape velocity at that radius will leave the cluster. If the mean thermal velocity exceeds the escape velocity, the resulting bulk outflow is termed {\it hydrodynamic escape}. {\it Tidal escape} can occur when material flows over the boundary between regions dominated by the cluster gravitational potential and the Galactic gravitational potential, which can be considered analogous to the cluster filling its Roche lobe.

If Jeans escape is to occur, the mean free path ($\lambda_{\rm MFP}$) of particles must be greater than the density scale length of the cloud at a point within the tidal radius. A discussion on the mean free path of particles can be found in many elementary textbooks (e.g.\ \citealt{FKR02}). It can be approximated by:
\begin{equation}
\lambda_{\rm MFP} = \frac{(4\pi\epsilon_0)^2 m^2 v^4}{2 \pi n q^4 \ln \Lambda} ,
\end{equation}
for particle mass $m$, velocity $v$, number density $n$ and charge $q$. For electron--ion interactions, the relevant masses and velocities are those of electrons, such that the mean free path is identical for electrons and ions. The Coulomb parameter, $\Lambda$, can be written in terms of the Debye length as $\Lambda = 4 \pi n \lambda_{\rm D}^3$, where:
\begin{equation}
\lambda_{\rm D} = \sqrt{\frac{\epsilon_0 k_{\rm B} T}{n q^2}} .
\end{equation}

Under Jeans-escape conditions, the particle-loss rate per unit time from the cluster becomes:
\begin{equation}
\dot{N}_{\rm Jeans} = 4 \pi r^2 \frac{n_{\rm e} v_{\rm th}}{2 \sqrt{\pi}} \left(1+\lambda_{\rm esc}\right) e^{-\lambda_{\rm esc}} ,
\label{JeansEq}
\end{equation}
where:
\begin{equation}
\lambda_{\rm esc}  = (v_{\rm esc} / v_{\rm th})^2
\label{lambdaEq}
\end{equation}
for escape velocity $v_{\rm esc}$. The thermal velocity is given by $v_{\rm th} = \sqrt{2 k_{\rm B} T / m}$. The bulk outflow velocity can then be derived simply as:
\begin{equation}
v_{\rm out} = \frac{\dot{N}_{\rm Jeans}}{4 \pi r^2 n} .
\label{JeansvEq}
\end{equation}
Note that Eqs.\ (\ref{JeansEq}) \& (\ref{JeansvEq}) implicitly assume that the velocity of the gas can be approximated by a Maxwell--Boltzmann distribution for the gas. This is not always met in plasma conditions, with particles of a few $kT$ being scattered to lower and higher energies (e.g.\ \citealt{NDS12}).

The mean-free path in a plasma is very short, due to the long-range electromagnetic interactions that can happen between ions and electrons. For the ICM densities here, it can typically be measured in AU or (at most) fractions of a parsec: several orders of magnitude smaller than the $e$-folding length at the tidal radius. Were a globular cluster isolated, Jeans escape would not be possible until many hundreds or thousands of parsecs from its core. We would naively expect tidal escape to dominate, however we will show that both can be important.

The physics of tidal escape are complex, and an accurate solution requires a full hydrodynamic model, in which the thermally expanding system is augmented by the kinematic and radiative acceleration of ICM by sources within the cluster and by processes around the $L_1$ and $L_2$ points. However, we can simplify the solution by analogy to stellar Roche lobe overflow (RLOF) of an eccentric binary, since the physics of RLOF is well studied (e.g.\ \citealt{Hilditch01,FKR02,IPS02}). The primary difference in our case is that the thermal timescale is much shorter than the dynamical timescale, the converse of the stellar case. Due to the large mass differential and large separation between the cluster and Galaxy, the Roche lobe for a globular cluster can be considered to be almost perfectly spherical, corresponding to the tidal radius. Meanwhile, the Roche lobe for the Galaxy is an almost-perfect plane which touches the cluster at its tidal radius. Since the cluster is orbiting around the Galaxy, we must consider both the $L_1$ and $L_2$ Lagrangian points.

Material reaching $L_1$ and $L_2$ will have no net attraction to the cluster, and will be free to stream away unhindered. The unhindered flow of material through the $L_1$ point can be simplified to:
\begin{equation}
\dot{M}_{\rm tidal} \approx \rho(r_{\rm t}) v S ,
\label{TidalEq}
\end{equation}
where the outflow velocity is approximately the sound speed ($v \sim c_{\rm s} \approx \sqrt{5/3\,kT/m_{\rm H}}$), and $S$ is the surface area over which the flow occurs. $S$ depends on the fraction by which the outer radius of the cluster ICM ($r_{\rm out}$) exceeds the tidal radius ($r_{\rm t}$). For the simplified situation where a spherical globular cluster potential interacts with a planar potential from the Galactic Plane, it can be simplified to be the surface area of a section through a sphere of radius $r_{\rm out}$ at a distance $r_{\rm t}$ from the cluster centre, hence:
\begin{equation}
\left(\frac{S}{r_{\rm t}}\right)^2 = \pi \left(\frac{r_{\rm out}}{r_{\rm t}}\right)^2 - 1 .
\end{equation}
While initially linear with fractional extent beyond the tidal radius, this quickly diverges to large values for $r_{\rm out} \gtrsim \sqrt{2} r_{\rm t}$.

Over the rest of the cluster, the absence of material above this radius means that the inward gas pressure will be removed (Eq.\ (\ref{PGEq})). We can expect the density and scale length will drop rapidly compared to Eq.\ (\ref{DensityEq}), as in planetary exospheres. The Jeans-escape criterion can then be satisfied. Material can then either leave the cluster via Jeans escape, or via flow through the $L_1$ and $L_2$ points. In either case, the escape is driven by thermal expansion of the ICM, and the response of the density structure to the escaping gas should then occur on the dynamic timescale of the cluster. We can therefore expect that \emph{the total ICM mass found within a cluster will be approximately equal to the mass lost by its stars over the dynamical timescale}. We examine whether this holds for 47 Tuc in Section \ref{FinalDensitySect}.

If it can be shown that the combination of Jeans- and tidal-escape rates from the cluster are similar to the mass-injection rate by stars, we can consider the modelled ionizing sources (post-AGB stars and white dwarfs) as capable of clearing mass from globular clusters through simple thermal expansion of the ICM.

In this work, we explore the role played by ionizing radiation in the cluster from more-evolved objects. In Section \ref{KnownSect}, we explore existing observations of ionizing sources within 47 Tuc. In Section \ref{MESASect} we model  the ionizing flux within 47 Tuc and M15 using the {\sc mesa} stellar evolution code, extending the results by analogy to other globular clusters. In Section \ref{CloudySect}, we explore a pressure-supported ICM model, and use {\sc cloudy} to investigate the temperature, density and ionization structure of the ICM in 47 Tuc. In Section \ref{StableSect}, we discuss the limit of cluster parameters over which an ionized ICM can be maintained, and conclude our findings in Section \ref{ConcSect}.


\section{Ionizing 47 Tucanae from known sources}
\label{KnownSect}

The cluster 47 Tucanae (NGC 104) offers the most stringent limits on the clearing of ICM thanks to careful studies of both its giant stars and its pulsars. \citet{FKL+01} find a correlation between pulse dispersion measures and radial acceleration among the cluster's pulsars, inferring an electron density in the cluster core of $n_{\rm e} = 0.067 \pm 0.015$ cm$^{-3}$. The pulsars cover a cylinder of radius 1.9 pc and depth 6.6 pc which can be homogenized to a sphere of radius 2.5 pc. Assuming these electrons come solely from ionized hydrogen and that the majority of the ICM is ionized, this can be converted to an equivalent mass of 0.1 M$_\odot$. At present, there is no evidence from pulsars that the distribution of electrons within the cluster core departs from a homogeneous cloud (P.~Freire, private communication). We can expect the ICM mass for entire cluster to be considerably higher.

\subsection{The minimum ionization rate}
\label{MinIonRate}

For a source to radiatively ionize the intracluster medium, it must satisfy the following criterion: the rate at which ionizing radiation is absorbed must be greater than the sum of the recombination rate in the ICM and the input of neutral material by stars. 

{\it Recombination rate:} Assuming a gas temperature of order 10\,000 K, the recombination rate of hydrogen in the cluster can be given as (e.g.\ \citealt{Harwit06}):
\begin{equation}
\Re_{\rm H} \approx 3 \times 10^{-13} n_{\rm e} n_{\rm p} \ \ [{\rm cm}^{3} {\rm s}^{-1}] ,
\label{RecombEq}
\end{equation}
where the electron and proton ($n_{\rm p}$) densities are approximately equal. Over the 2.5-pc central region of 47 Tuc, this equates to a recombination rate of 2.6 $\times$ 10$^{42}$ atoms s$^{-1}$. A surface brightness of $10^{-23}$ W m$^{-2}$ arcsec$^{-2}$ would be seen in H$\alpha$ at Earth, assuming that each recombination gives rise to a Balmer-$\alpha$ photon. If a 50-\AA\ H$\alpha$ filter is employed, this equates to $\sim$27.5 mag arcsec$^{-2}$. This would be impossible to detect from Earth, especially given the confusion from light emitted by the cluster's stars.

{\it Neutral mass input by stars:} A stellar death rate of one star per 80\,000 years \citep{MBvLZ11}, and an average initial and final stellar mass of 0.87 and 0.53 M$_\odot$, respectively (\citealt{RFI+97,MKZ+04,KSDR+09,MBvL+11,Kalirai13}; McDonald et al., in prep.), implies that the cluster's stars exhibit a total mass-loss rate\footnote{We revise these numbers based on stellar models later: see Table \ref{ClusterTable}.} of $\dot{M}_\ast \approx 4 \times 10^{-6}$ M$_\odot$ yr$^{-1}$ or $\dot{N}_\ast \approx 1.6 \times 10^{44}$ amu s$^{-1}$. The majority of this should come as cool, slow winds from the cluster's giant stars\footnote{ Our {\sc mesa} model, described later in the text, indicates $\sim$71 per cent of the mass loss occurs on the RGB and AGB above 700 L$_\odot$, where winds are expected to have an outflow velocity of $\sim$10 km s$^{-1}$ \citep{MvL07,MAD09,Groenewegen14} and be cool enough to start forming molecules and dust \citep{MBvL+11,MBvLZ11}. Lower-luminosity stars may have faster, hotter winds (e.g.\ \citealt{DSS09}) but do not contribute much to the total mass-loss rate.}. Assuming all this material must be ionized, $\gtrsim$10$^{44}$ ionizing photons s$^{-1}$ must be absorbed to maintain the ionization of the ICM.

We will later show that the inner 2.5 pc accounts for $\sim$5.7 per cent of the cluster mass (based on a Plummer model applied in Section \ref{Model47Tuc}), hence the mass input in the inner 2.5 pc should be around $9 \times 10^{42}$ amu s$^{-1}$. In this region, the mass-injection rate can therefore be expected to dominate over the recombination. To maintain ionization, $\gtrsim$10$^{43}$ photons s$^{-1}$ must be absorbed in the inner 2.5 pc.

Recombination rates scale with the square of the electron density (strictly $n_{\rm e} n_{\rm p}$). However, the mass-injection rate by stars scales with the stellar density. As both electron and stellar densities fall off with radius from the cluster centre, we can expect the creation of neutral matter to become even more dominated by mass input by stars as we move further from the cluster core. Ionization of the mass injected by all stars within the cluster therefore requires that $>$1.6 $\times$ 10$^{44}$ photons s$^{-1}$ with wavelengths $<$912 \AA\ are absorbed by the ICM, which equates to $>$2 $\times 10^{26}$ W or $>$0.6 L$_\odot$ of ionizing radiation.

\subsection{High-energy non-thermal sources}
\label{HighESect}

\begin{figure}
\centerline{\includegraphics[height=0.47\textwidth,angle=-90]{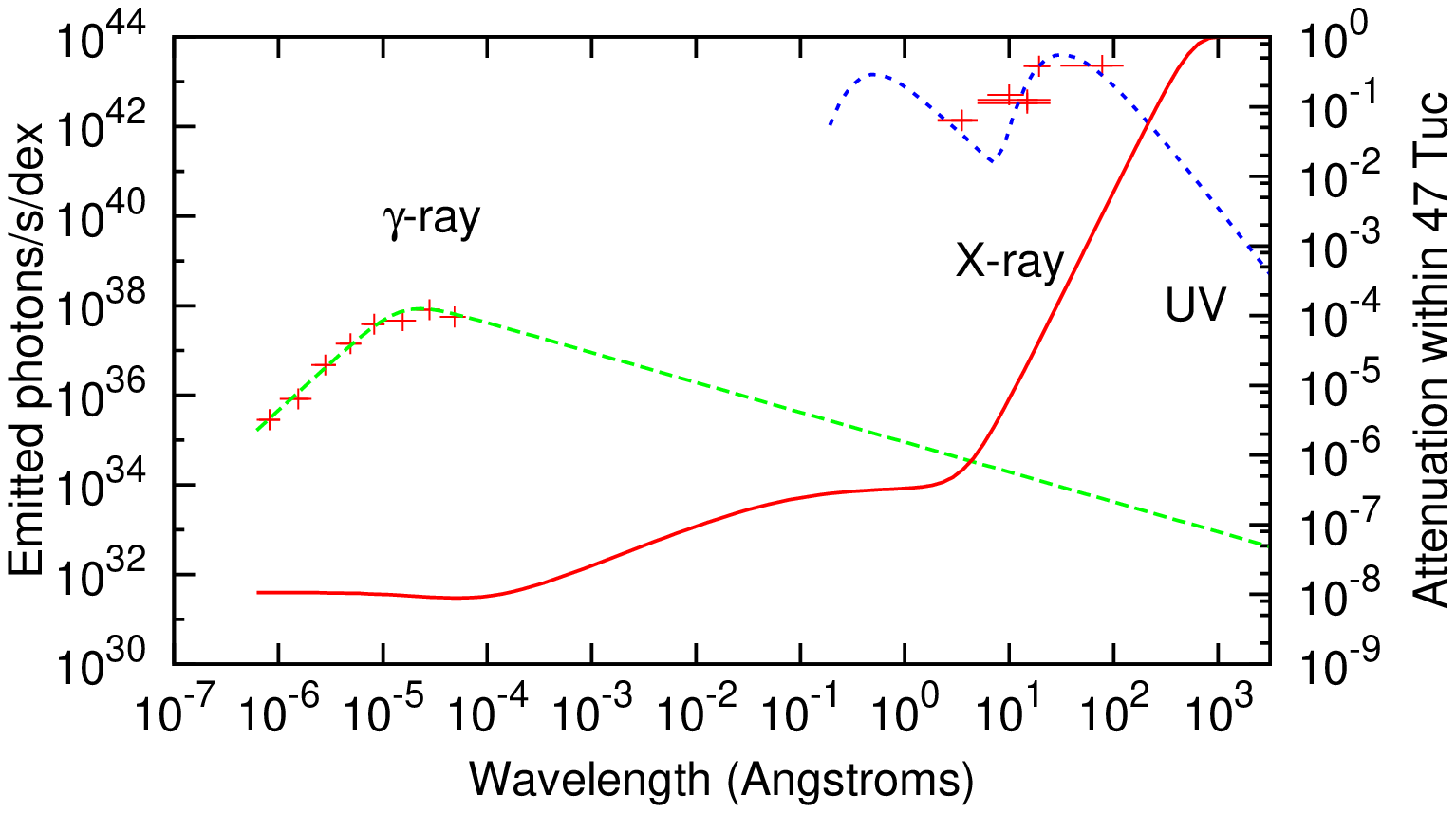}}
\centerline{\includegraphics[height=0.47\textwidth,angle=-90]{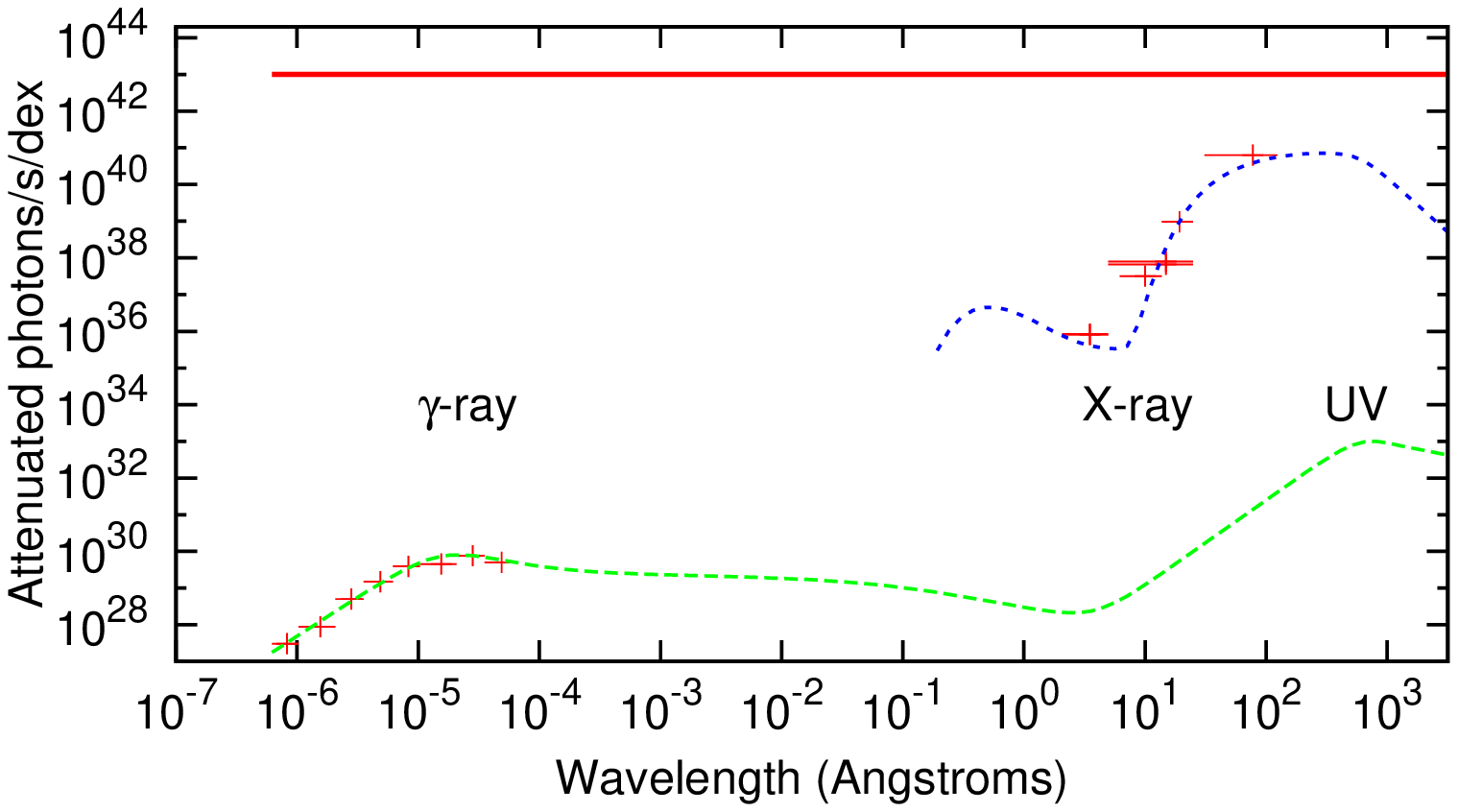}}
\caption{The radiation environment of 47 Tucanae, as observed by \emph{Fermi}, \emph{Chandra}, \emph{EUVE} and \emph{ROSAT} (red points), normalized to the number of photons produced in a band covering a factor of ten in wavelength. \emph{Top panel:} The green (dashed) and blue (dotted) curves show our models of the emission in the $\gamma$- and X-ray components. The red (solid) curve shows the fraction of those photons which would be absorbed by the neutral hydrogen medium within 2.5 pc of the centre of 47 Tucanae. \emph{Bottom panel:} The number of absorbed photons for each component (the product of emitted photons times attenuation), which never exceeds the limit required to maintain ionization of all the neutral medium in the inner 2.5 pc (red line, $\sim$10$^{43}$ photons s$^{-1}$).}
\label{RadFig}
\end{figure}

The observed high-energy spectral energy distribution (SED) of 47 Tuc is shown in Figure \ref{RadFig}, and described in detail below. To create this figure, we correct the observed fluxes for attenuation by Galactic interstellar hydrogen\footnote{Attenuation co-efficients are taken from the National Bureau of Standards report NSRDS-NSB 29: \tt{http://www.nist.gov/data/nsrds/NSRDS-NBS29.pdf; http://physics.nist.gov/PhysRefData/XrayMassCoef/ElemTab/z01.html}} ($N_{\rm H} = 10^{20.11}$ cm$^{-2}$; \citealt{HGE+05}) and for interstellar dust ($E(B-V) = 0.04$ mag; \citealt{Harris10}), then use a distance of 4.5 kpc to 47 Tuc to determine the number of ionizing photons at the source. We then compute the attenuation due to the column density of hydrogen within the inner 2.5-pc region ($N_{\rm H} = 10^{17.71}$ cm$^{-2}$). We assume a spherically symmetric flux output from the cluster. Gaps in the observed spectral coverage of 47 Tucanae mean that we can only approximate both the number of ionizing photons within the cluster, and what fraction is absorbed.

\subsubsection{$\gamma$-ray pulsars}

Globular cluster $\gamma$-ray emission arises primarily from synchrotron radiation from pulsar winds. Pulsars are thought to lose a substantial fraction (1--100 per cent) of their spin-down lumionsity through high-energy radiation (e.g.\ \citealt{Petri12,Caraveo13}). Observations of 47 Tuc have been made in the 0.2--20 GeV ($6.2 \times 10^{-7}$ -- $6.2 \times 10^{-5}$ \AA) range by \emph{Fermi}, finding a turnover in the energy spectrum around 2.2 GeV ($5.6 \times 10^{-6}$ \AA; \citealt{AAA+10}). We use a simple double-power-law model to model the $\gamma$-ray emission from the cluster, finding (for wavelength $\lambda$ in metres):
\begin{eqnarray}
\frac{d N_\gamma}{d\lambda} \!\!\!\!&\approx&\!\!\!\! 9 \!\times\! 10^{34} \ {\rm s}^{-1} {\rm dex}^{-1} \!\left(\frac{\lambda}{10^{-10}}\right)^{\!\!\!\!-\frac{2}{3}} \!\!\left(1+ \frac{1.5 \!\times\! 10^{-15}}{\lambda}\right)^{\!\!\!\!-2.8} .\nonumber \\
\ 
\label{GammaEq}
\end{eqnarray}
This reproduces the turnover found by \citet{AAA+10} (see Figure \ref{RadFig}). Integrating Eq.\ (\ref{GammaEq}) over wavelength, we find that the total number of $\gamma$-ray photons is $N_{\gamma} \approx 9 \times 10^{39}$ photons s$^{-1}$. Attentuation co-efficients here are small, at $\approx$1 part per billion (ppb) within the cluster core (column densities and attenuation co-efficients are given at the start of Section \ref{HighESect}). Only $\approx$9 $\times$ 10$^{31}$ photo-ionizations s$^{-1}$ are expected by $\gamma$-ray photons, far short of the $\sim$10$^{43}$ required in the inner 2.5 pc. The long mean free path of photons and the weak attenuation mean that we discard secondary events resulting from these very-high-energy photo-ionizations, thus $\gamma$ rays do not contribute significantly to the ionization of material in the ICM.

An alternative approach would be to look at the pulsars themselves. The spin-down luminosity of a pulsar is given by \cite{pulsarhandbook}: 
\begin{equation}
L = \frac{4 \pi^2 \dot{P} \ \cdot \ \frac{2}{5}MR^2} {P^3} ,
\end{equation}
where $M$, $R$, $P$ and $\dot{P}$ signify the mass, radius, period and period derivative of the pulsar, respectively. In clusters, $\dot{P}$ must be corrected due to the gravitational acceleration of the pulsar towards the cluster potential, resulting in some uncertainty in addition to considerable errors in their masses and radii. In comparison to the otherwise well-determined parameters of pulsars, the spin-down luminosity retains a considerable fractional uncertainty. The spin-down lumionsity of the known pulsars in 47 Tuc\footnote{http://www.naic.edu/~pfreire/GCpsr.html} varies by a factor of $\sim$30 \citep{FCL+01}. Assuming each pulsar has a mass of 1.4 M$_\odot$, and is 10 km in radius, this corresponds to a spin-down luminosity of 10$^{26}$ -- 10$^{27.5}$ W per pulsar \citep{GCH+02}. If the $\sim$20 observed pulsars are taken from a population of $\sim$1000 (cf.\ \citealt{Meylan88}), and 1 ppb of their radiation is absorbed, this equates to $\sim$10$^{21}$ W of absorbed radiation: much less than the minimum 2 $\times 10^{26}$ W required to ionize the ICM.


It is possible that the ICM is also heated and ionized by thermalisation of relativistic ions ejected by pulsars. This can occur via inverse Compton scattering, where the relativistic ejecta interacts with the low-density ICM plasma, and should manifest itself as diffuse, high-energy radiation within the cluster. A diffuse, X-ray component has been suggested from this mechanism, but the flux ($\approx$2 $\times$ 10$^{25}$ W; \citealt{WHK+14}) is only a tenth of that required to ionize the ICM. We do not consider pulsars as sufficiently effective radiative or kinematic heaters of the ICM.



\subsubsection{X-ray and extreme UV sources}

X-ray and extreme UV sources in 47 Tuc have been measured by \emph{Chandra} \citep{HGE+05} and by both the \emph{R\"ontgensatellit (ROSAT)} All-Sky survey and pointed Poisition Sensitive Proportional Counter (PSPC) observations \citep{VH98}. The cluster was not detected by the \emph{Extreme Ultra-Violet Explorer (EUVE;} 70--760 \AA\emph{)}, nor in the \emph{ROSAT} Wide Field Camera (60--300 \AA) images. This may indicate that the X-ray emission is likely to peak around 50\AA\ (certainly no sources have been found that should have important contributions longward in these wavelengths), however the intra-cluster and Galactic hydrogen can be expected to become opaque at around these wavelengths.

The \emph{Chandra} data show that X-ray emission from the cluster is dominated by mass transfer systems: quiescient low-mass X-ray binaries (qLMXBs; typically main-sequence + neutron stars) and cataclysmic variables (CVs), plus a less-significant contribution from X-ray-active binary stars. At least the brightest sources are considerably time-variant. Milli-second pulsars only account for a very minor component ($\lesssim$1 per cent) of the X-ray flux, tending to emit softer X-rays than the CVs, which dominate at higher energies.

The SED of the qLMXBs (and by extension the CVs) can be modelled as a blackbody peaking at a few keV, which then undergoes Comptonisation in the hot plasma \citep{GPO01}. The blackbody could be emitted either from the surface of the compact object, the inner edge of the accretion disc, or an accretion disc hot spot, and therefore may include contributions from plasmas of a variety of different temperatures. Typically, the spectrum steepens considerably above $\approx$15 keV ($\approx$1\AA), such that negligible emission occurs at higher energies. This is fortunate, as a gap in spectral coverage exists from 6 keV to 200 MeV, directly between the tail of the X-ray emission and the peak of the pulsar $\gamma$-ray emission, which we presume is largely devoid of flux.

In terms of photon counting, the thermal component dominates over the Compton component, even though most of the energy can be in the Comptonized component \citep{GPO01}. The mass absorption co-efficient at higher energies (above a few keV) is considerably less than that experienced by the thermal blackbody, and it plateaus at around 0.45 cm$^2$ g$^{-1}$ over the 7--70 keV region appropriate for Compton scattering \citep{Hubbell71}. We can therefore express the X-ray flux as a blackbody, of which a certain fraction of photons ($\approx$15--80 per cent) will be upscattered and experience the lower (0.45 cm$^2$ g$^{-1}$) mass absorption co-efficient.

We estimate the X-ray emission from sources within the cluster using the unabsorbed X-ray fluxes and $kT$ values listed by \citet[][their table 7]{HGE+05}. \citet{HGE+05} model the hydrogen column density to be substantially higher than the intervening Galactic column for several sources. To be conservative, we use their unabsorbed values, negating any intrinsic absorption in the source.

Summing the brightest X-ray sources detected by \emph{Chandra}, we find a total ionizing flux (integrated across $0 \leq \lambda \leq 912$\AA) of $\approx$4 $\times 10^{41}$ photons s$^{-1}$, of which $\sim$9 $\times 10^{31}$ photons s$^{-1}$ will be absorbed: 11 orders of magnitude too few. Hence, we do not consider X-rays to be a likely ionization source for intercluster hydrogen.

\subsection{Thermal UV sources}
\label{UVSect}

\subsubsection{Modelling the data}



\begin{figure}
\centerline{\includegraphics[height=0.47\textwidth,angle=-90]{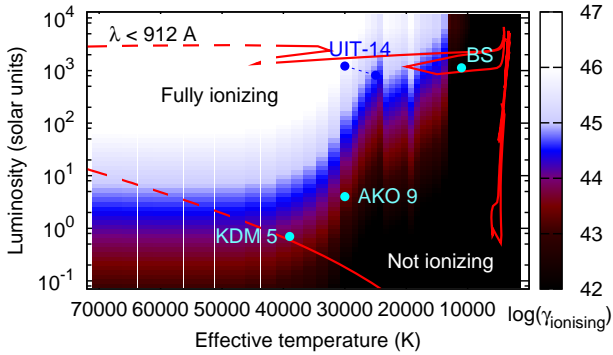}}
\caption{Number of ionizing photons ($\lambda < 912$\AA) produced by a star of given temperature and luminosity, computed using {\sc BT-Settl} model atmospheres. A flux of $\sim$10$^{44}$ photons is required to ionize the ICM of 47 Tuc: any objects in the white region can fully ionize the entire ICM on their own. Individual objects in the black region will not contribute significantly to the ionization. Known objects within 47 Tuc are shown, as is the {\sc mesa} stellar evolution track we model in this work (red line). KDM 5 has a predicted white dwarf age of $\sim$3 Myr.}
\label{IonizeFig}
\end{figure}

Attenuation in the range $\sim$50 to 912 \AA\ is sufficient that a neutral hydrogen medium in the core of 47 Tuc should absorb any photons emitted at these wavelengths. For a stellar source, the number of ionizing photons is simply the fraction of the stellar flux ($F$) produced shortward of the ionization wavelength ($\lambda_{\rm i} = 912 \AA$ for H {\sc i}), divided by the energy per photon ($hc/\lambda$) and normalised by the total number of photons ($\gamma_{\rm total}$): 
\begin{equation}
\gamma_{\rm ionizing} = \gamma_{\rm total} \frac{\int_0^{\lambda_{\rm i}} F_\lambda \lambda / (h c) \ d\lambda}{\int_0^\infty F_\lambda \lambda / (h c) \  d\lambda} ,
\label{IonEq}
\end{equation}
The total number of photons from a blackbody is given by (e.g.\ \citealt{MW12}):
\begin{equation}
\gamma_{\rm total} = 1.5205 \times 10^{15} \times 4 \pi R^2 T_{\rm eff}^3 ,
\end{equation}
for radius $R$ and effective temperature $T_{\rm eff}$, which can re-written in luminosity terms using $L = 4 \pi R^2 \sigma T^4$:
\begin{equation}
\gamma_{\rm total} = 1.5205 \times 10^{15} \times \frac{L}{\sigma T_{\rm eff}} .
\end{equation}
We calculate the fraction of ionizing photons in Eq.\ (\ref{IonEq}) using a grid of {\sc BT-Settl} model atmospheres\footnote{We use {\sc BT-Settl} atmospheres as they closely reproduce the observed flux in the $\approx$1000\AA\ region (see \citealt{DC13}).}, to identify which sources are able to produce the $\sim$10$^{44}$ photons required to ionize the ICM. The results are shown in Figure \ref{IonizeFig}. We find that the effective temperature of the source is critical: an increase of a few tens of percent in temperature (at constant luminosity) can lead to a factor of ten increase in ionizing flux. Any post-AGB star above $T_{\rm eff} \gtrsim 14\,000$ K and any white dwarf above $L \gtrsim 1$ L$_\odot$ can generate enough radiation to ionize the entire cluster's intra-cluster hydrogen.

\subsubsection{Individual sources}

\begin{figure}
\centerline{\includegraphics[height=0.47\textwidth,angle=-90]{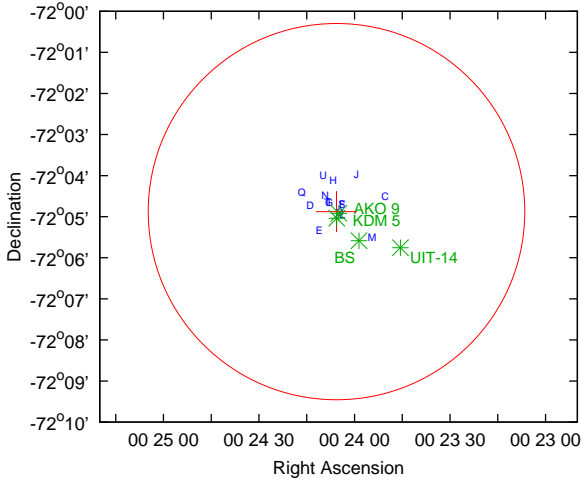}}
\caption{Location of important sources within the cluster. Pulsars are labelled by their letter, in blue. Bright UV sources are denoted by green asterisks and labelled. The red cross marks the cluster centre, and the outer circle marks the Plummer scale radius of $a = 5.98$ pc.}
\label{MapFig}
\end{figure}

Observationally, the far-UV ($\approx$1500\AA) flux from 47 Tuc is strongly dominated by one post-AGB star (\citealt{OCDS+97,SDS+12}; the ``bright star'' in their \emph{GALEX} and \emph{Ultraviolet Imaging Telescope (UIT) data}). Spatially incomplete samples of sources found in \emph{Hubble Space Telescope} data show that the majority of the remaining UV flux comes from probable post-AGB stars, white dwarfs and the CV AKO 9 \citep{KDMA+08,WGK+12}. However, the $F_{912} - F_{1500}$ colours of these stars are significant: most sources observed to be bright at 1500 \AA\ will have very red $F_{912} - F_{1500}$ colours, hence produce negligible ionizing radiation. We list several important sources below and show them in Figures \ref{IonizeFig} and \ref{MapFig}.

\emph{Bright (post-AGB) star:} \citet{DDF95} quote atmospheric parameters for this B8 giant star of $T_{\rm eff} = 11000$ K, $\log g = 2.5$ and $L = 1125$ L$_\odot$ (corrected to a distance of 4.5 kpc). This star should provide about $4 \times 10^{41}$ ionizing photons. This is not enough to have a significant effect on the intra-cluster hydrogen.


\begin{figure*}
\centerline{\includegraphics[height=0.95\textwidth,angle=-90]{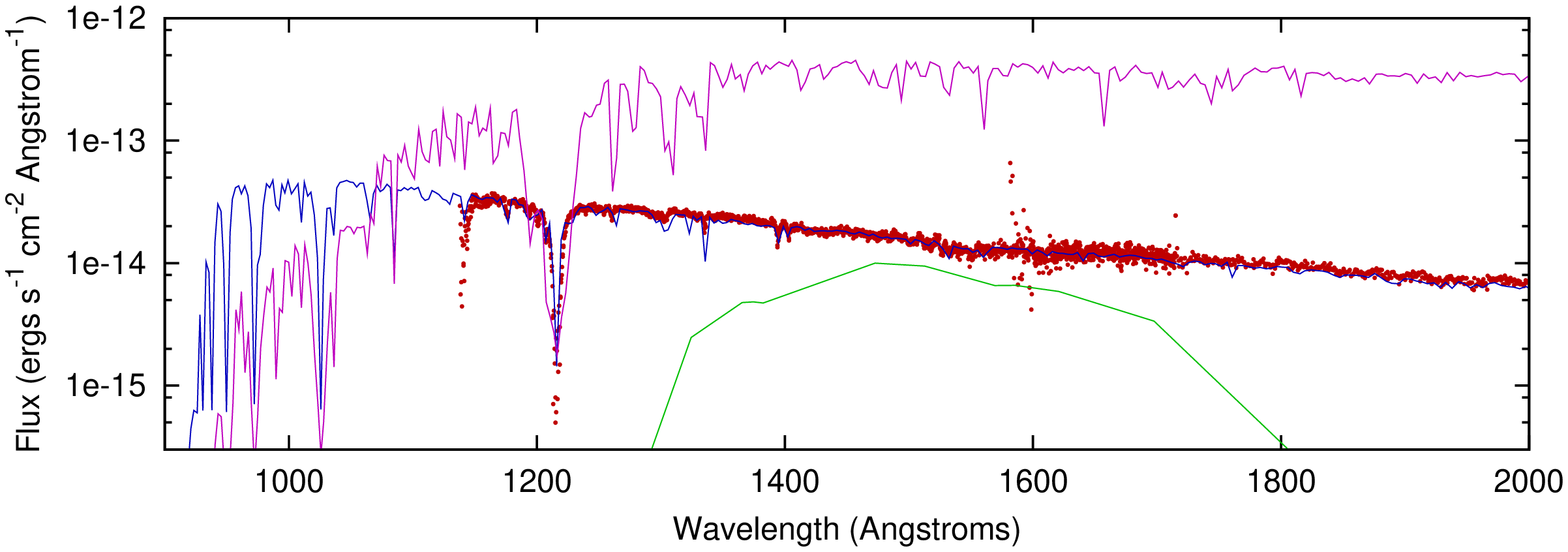}}
\centerline{\includegraphics[height=0.95\textwidth,angle=-90]{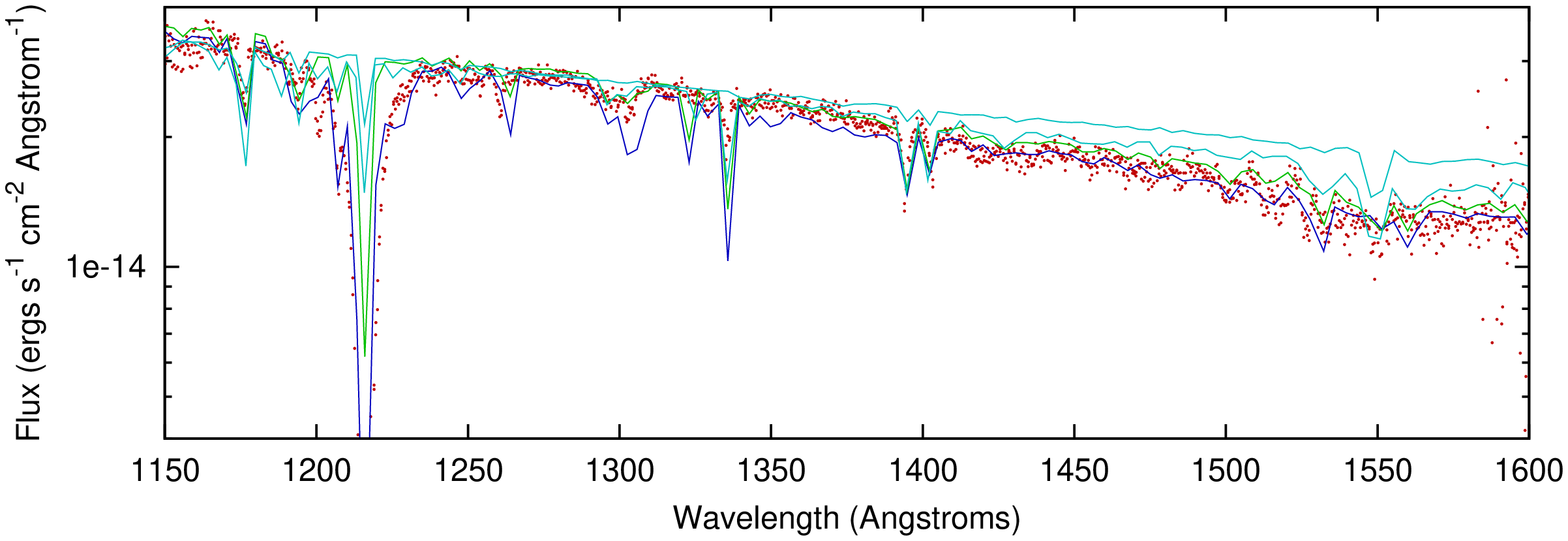}}
\caption{\emph{Hubble Space Telescope} STIS spectrum of UIT-14. Top panel: red points show the archival STIS spectrum, the lower, green line shows the \emph{GALEX} FUV transmission curve. The middle, blue curve shows a 25000 K, $\log g$ = 4.5 {\sc BT-Settl} atmosphere model, fit to the spectrum. The upper curve shows a 11200 K, $\log g = 2.5$ model, scaled to the flux of the UV-bright post-AGB star. Note that the intrinsically fainter UIT-14 dominates below the 1120 \AA\ CO dissociation wavelength. Bottom panel: close-up of the spectrum of UIT-14, showing the same {\sc BT-Settl} model in blue. The upper (green and cyan) lines show models at 30000 K, with $\log g$ of 3.0, 4.5 and 6.0, which poorly reproduce the Lyman $\alpha$ line at 1216 \AA\ or the spectral shape and line complex around 1550 \AA. }
\label{UITFig}
\end{figure*}

\emph{UIT-14} (00$^h$23$^m$45.59$^s$, --72$^\circ$05$^\prime$45.3$^{\prime\prime}$): the \emph{UIT} data of \citet{OCDS+97} suggests that UIT-14 would be a significant far-UV source within the cluster, with $T_{\rm eff} \approx 50000$ K. However, a cooler temperature is required to fit both the observed \emph{UIT} and archival \emph{Hubble Space Telescope (HST)} spectra. A simple `by-eye' fit to the spectrum suggests its temperature is only around $T_{\rm eff} \approx 25000$ K, with $\log g \approx 4.5$, with an implied luminosity of $\approx$400 L$_\odot$. The surface gravity cannot be much below or above this in order to reproduce the shape and depth of the Lyman $\alpha$ line, respectively. Similarly, the temperature cannot be altered by more than a few thousand degrees to avoid removing the finer spectral features, or making them too strong, particualrly in the 1300--1700 \AA\ region.

Using these parameters, UIT-14 should indeed dominate the flux output at $\lambda \lesssim 1070$ \AA, producing $1 \times 10^{44}$ photons at wavelengths below the Lyman limit. Increasing the temperature to 30\,000 K, however, produces $5 \times 10^{45}$ sub-Lyman photons. Thus, UIT-14 alone may ionize most or all of the cluster ICM.

\emph{AKO 9:} \citet{MMP+97} and \citet{KZS+03} report on the UV-bright CV AKO 9. \citet{KDMA+08} model the mass-accreting white dwarf as a 30\,000 K, 0.07 R$_\odot$ object. At a corresponding luminosity of 4 L$_\odot$, this star would produce around $3 \times 10^{44}$ photons at wavelengths below the Lyman limit. \citet{KDMA+08} note that AKO 9 appears to erupt on a characteristic timescale of $<$7 years. During outburst, the star's temperature and/or luminosity must increase, probably quite substantially (see their figure 9), meaning the ionizing flux from this star would likely also substantially increase.


\emph{Other stars:} \citet{KDMA+08} model a number of other objects in their \emph{HST} field which contribute significantly to the far-UV flux. Of these, the most significant is their star 5 (KDM 5 in Figure \ref{IonizeFig}): a 39\,000 K, 0.17 R$_\odot$, 0.7 L$_\odot$ white dwarf, which should produce around $8 \times 10^{43}$ photons at $\lambda < 912$ \AA\ and $1 \times 10^{44}$ photons at $\lambda < 1120$ \AA. Other sources they mention contribute roughly $1-2 \times 10^{44}$ and $3-5 \times 10^{44}$ photons, respectively.

Their field covers a third of the cluster's core, from which we presume the total UV flux from other sources within the globular must be about $\sim$6$\times$ these values, or $\sim$10$^{45}$ and $\sim$10$^{45.5}$ photons s$^{-1}$. Despite their brightness shortward of the Lyman $\alpha$ limit, most of these sources are too faint to be detected in the \emph{GALEX} FUV filter: due to the stark contrast in spectral output on different sides of the Lyman break, many other sources are sufficiently bright in the FUV to crowd out such hot white dwarfs.

These results would imply there is 10--50$\times$ the required flux to ionize the intracluster hydrogen in 47 Tuc. However, this is solely based on the observed objects. A single, unnoticed, hot ($\gtrsim$70\,000 K) white dwarf could easily increase the UV flux by another order of magnitude or more.


\subsection{The Str\"omgren radius of 47 Tuc from known sources}

It is clear that several sources provide the $\approx$10$^{44}$ ionizing photons s$^{-1}$ needed to maintain ionization of the ICM of 47 Tucanae and that, very approximately, their sum should be $\sim$10--50$\times$ this value (depending on the relative contribution of UIT-14). Assuming a total estimated ionizing flux from known sources in the region of $1-6 \times 10^{45}$ photons s$^{-1}$ and a recombination rate of 1.4 $\times$ 10$^{-15}$ cm$^{-3}$ s$^{-1}$, one obtains a Str\"omgren sphere radius of 110--200 pc. This easily exceeds the size of the cluster (without accounting for the drop in density from the central value) and means that the ionization front must be so far from the central region as to be in the surrounding, already-ionized Halo gas. We therefore expect no neutral hydrogen inside the cluster. Any injected material will be rapidly ionized, with post-AGB stars and hot white dwarfs as the dominant sources of ionization.

We note that we largely restrict our analysis in this work to hydrogen ionization. However, we also except helium ionization: radiation above the helium edge (504.3 \AA; \citealt{LW55}) will maintain helium ionization and prevent it from effectively cooling the ICM. This effect is likely to be larger in globular clusters than the field due to the contribution to the X-ray flux by post-AGB stars and white dwarfs, which have a harder X-ray flux than the typical Galactic interstellar radiation field \citep{KSM+14}. We discuss ionization of elements other than hydrogen in Section \ref{InitialDensitySect} and, in particular, Figure \ref{CloudyIonFig}.




\section{The long-term ionization of ICM}
\label{MESASect}

\subsection{Simulating 47 Tuc}

\subsubsection{The stellar evolution model}

\begin{figure}
\centerline{\includegraphics[height=0.47\textwidth,angle=-90]{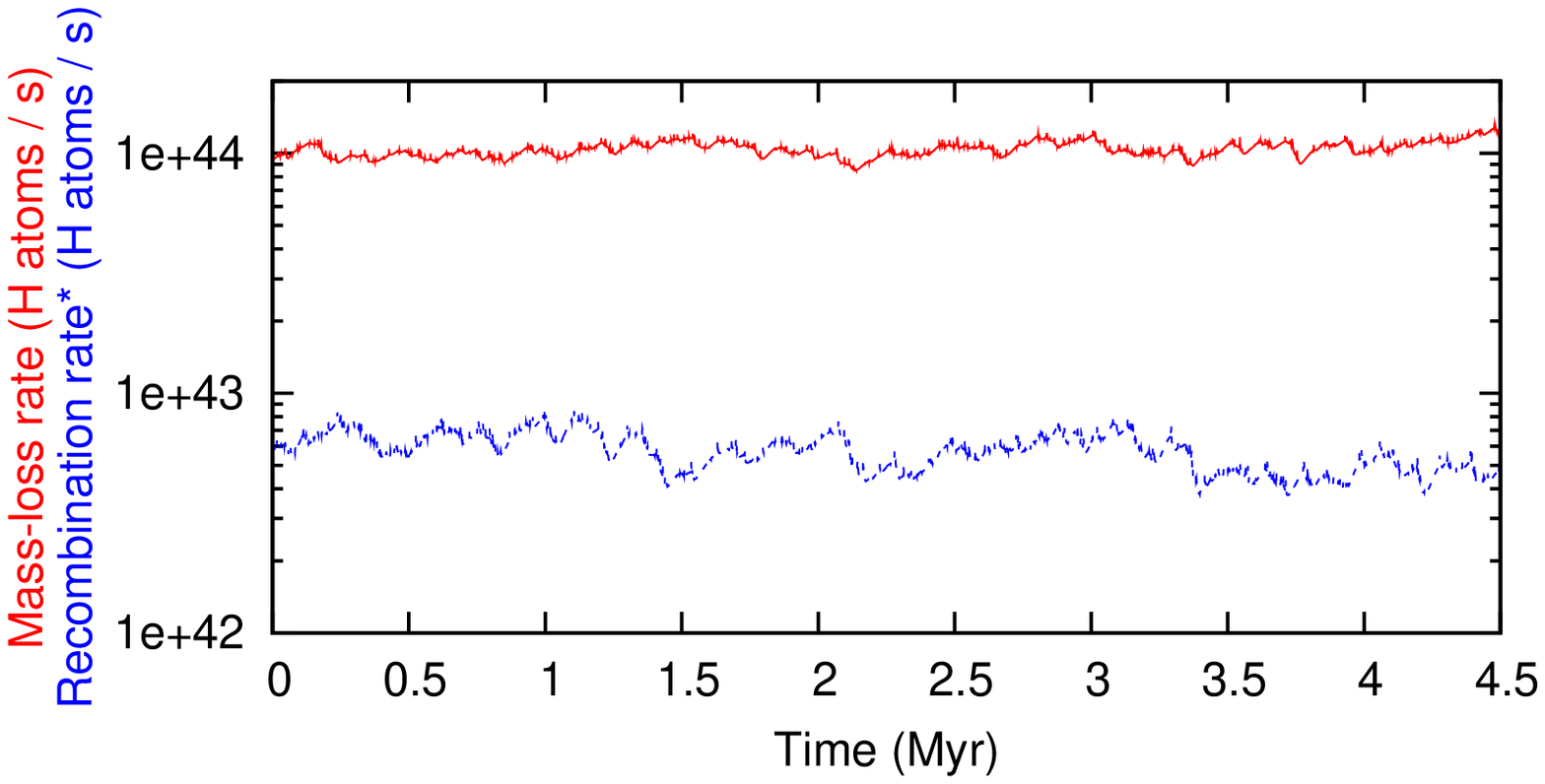}}
\centerline{\includegraphics[height=0.47\textwidth,angle=-90]{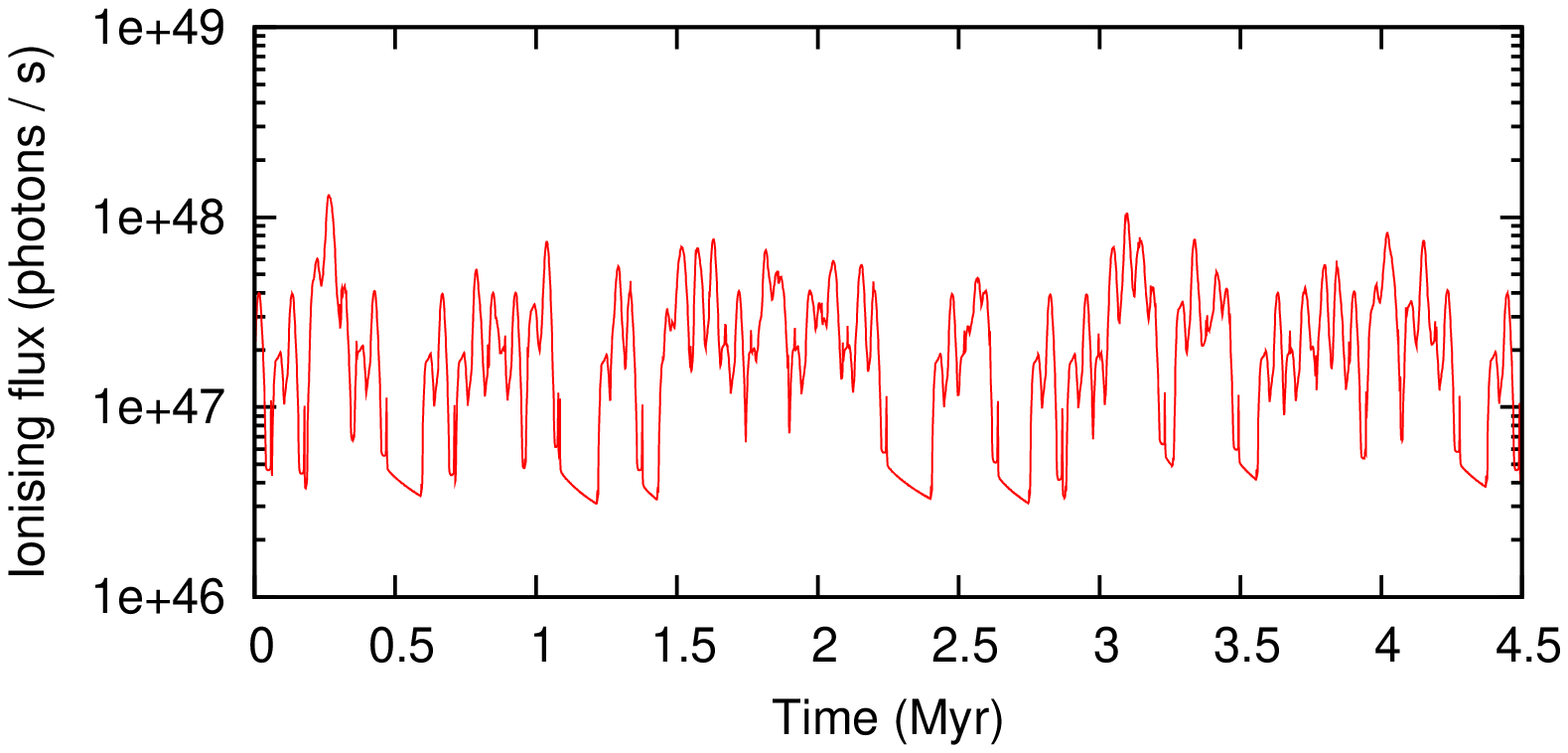}}
\centerline{\includegraphics[height=0.47\textwidth,angle=-90]{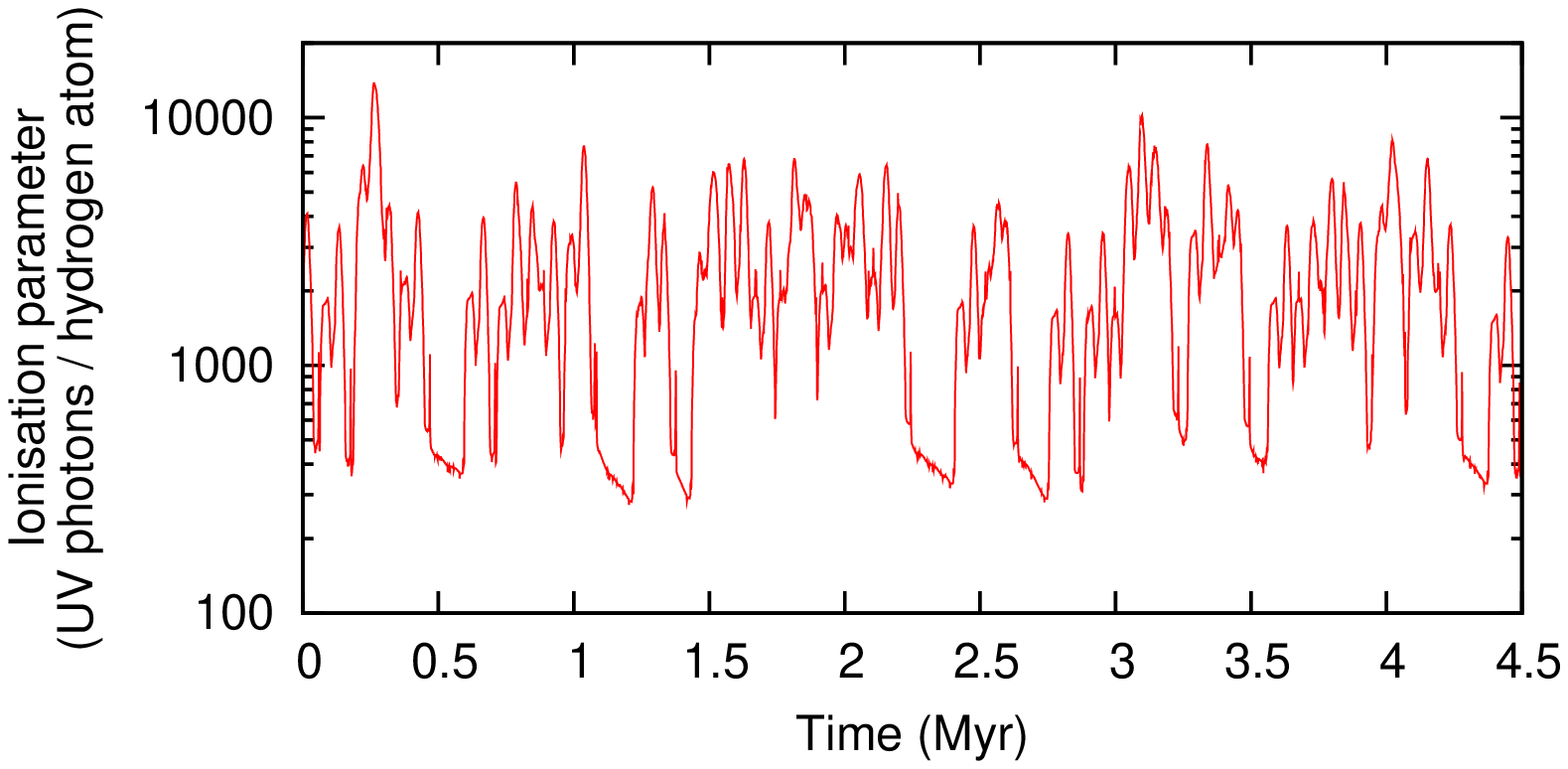}}
\centerline{\includegraphics[height=0.47\textwidth,angle=-90]{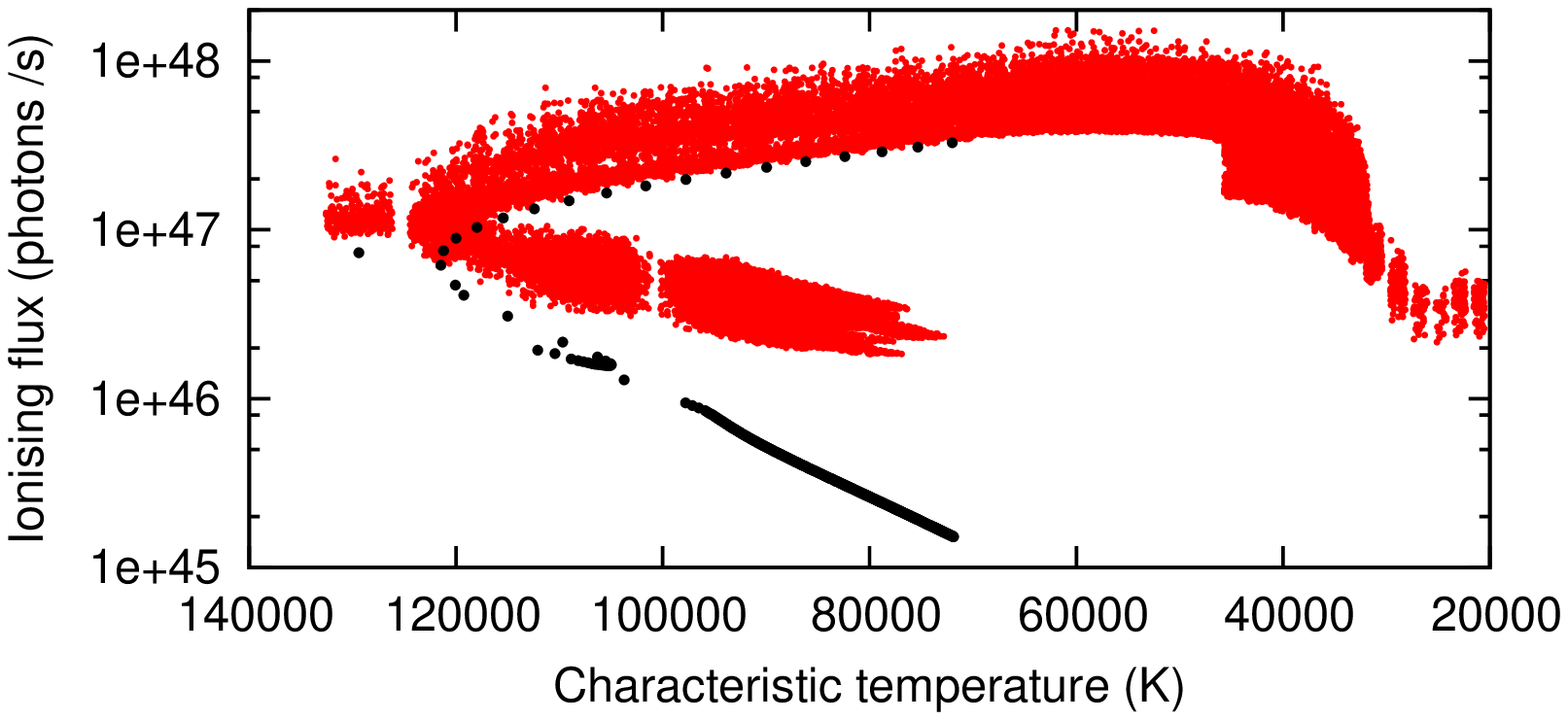}}
\caption{Top panel: total modelled mass-injection rate from RGB and AGB stars and the recombination rate of interstellar hydrogen atoms \emph{assuming that particle density scales with mass-injection rate}. Second panel: rate of ionizing photon production within the cluster. Third panel: ratio of ionizing photon rate to mass-injection rate. Bottom panel: in red (grey in print), the characteristic temperature of the ionizing radiation (a random spread of 5 per cent in temperature has been applied to smooth the distribution); in black, the output of the hottest source in each cluster, which is always $\gtrsim$70\,000 K.}
\label{IonizeFluxFig}
\end{figure}

To investigate the properties of ionizing stars in 47 Tuc, we created a stellar evolution model using the {\sc mesa} (Modules for Experiments in Stellar Astrophysics) code \citep{PBD+11,PCA+13}, appropriate for a post-AGB star in 47 Tucanae. The evolution of an (initially) 0.899 M$_\odot$ star was followed, setting the mass-loss rate following \citet{Reimers75}, with $\eta = 0.41$ (McDonald et al., submitted MNRAS\footnote{This value of $\eta$ is derived from comparing horizontal-branch star masses \citep{GCB+10} to initial masses from stellar isochrones. In our submitted work, we find that, typically, $\eta = 0.4$--0.5 in globular clusters. This compares to the slightly higher $\eta = 0.5$--0.6 found by \citet{SC05}, and simliar values from field stars \citep{CS11}. Conversely, astroseismological results from \emph{Kepler} observations of NGC 6791 suggest a slightly lower $\eta$.}). We adopted elemental abundances and atmospheric opacities appropriate for an [Fe/H] = --0.72 dex, [$\alpha$/Fe] = +0.3 dex star. This model reproduced the observed RGB tip at an age of 12 Gyr (\citealt{SW02,DAPC+05,MFAP+09,DSA+10,VBLC13}; McDonald et al., submitted), producing a horizontal branch star of mass 0.678 M$_\odot$ \citep{GCB+10}. The model underwent seven thermal pulses, including a late thermal pulse immediately after leaving the AGB and a very late thermal pulse near the maximum post-AGB temperature of 129\,400 K. The model ended with a cooling white dwarf of mass 0.546 M$_\odot\ \!^{[\!\,}$\footnote{Observed masses of white dwarfs in globular clusters are $\sim$0.53 M$_\odot$ \citep{RFI+97,MKZ+04,KSDR+09}. This follows the observed flattening of the initial--final mass relation at low masses (e.g.\ \citealt{KHK+08,GZHS14}).}$^{]}$.

We interpolated or averaged this model to a fixed temporal grid of 1000-year steps, obtaining the mass-loss rate, temperature, luminosity and gravity at each step. For each step, we took the eight nearest {\sc BT-Settl} model atmospheres and interpolated between them in metallicity, temperature and surface gravity to obtain a SED appropriate for the star. As in Section \ref{UVSect}, we calculated the number of photons emitted below the Lyman break (912 \AA) for each time step.

The maximum temperature of the underlying {\sc bt-settl} models is 70\,000 K, and the model star spends 557\,000 years above this temperature. In this phase, we divide the 70\,000 K {\sc bt-settl} model by a 70\,000 K blackbody and multiply by one of the appropriate temperature. This likely represents an under-estimate of the UV flux from a real star during this period as the Lyman break is less pronounced in the hottest stars. The maximum UV flux attained in the model is 3.6 $\times$ 10$^{47}$ ionizing photons s$^{-1}$.

We simulate 47 Tucanae by randomly recreating a stellar `death' on average every 80\,000 years, allowing a new RGB star to evolve through to the white dwarf phase. At every 1000-year timestep, we calculate the mass-injection rate by giant stars and the UV flux from post-AGB stars and white dwarfs. We assume the particles injected into the cluster are composed of a 75:25 (by mass) hydrogen:helium plasma which is fully ionized, hence the average particle mass is 7/9 amu. These rates and fluxes can simply be summed to determine the rate and flux for the entire cluster. These are presented in Figure \ref{IonizeFluxFig}.

\subsubsection{The cluster's long-term ionization}

Firstly, we notice that the mass-injection rate into 47 Tucanae is relatively constant, remaining near 1 $\times$ 10$^{44}$ hydrogen atoms per second. The timescale for mass loss near the tip of both the RGB and AGB is sufficiently longer than the 80\,000-year death timescale of the cluster. Assuming the intra-cluster ion density scales with the mass-injection rate, the recombination rate also remains relatively constant at a few $\times$ 10$^{42}$ recombinations per second. The recombination rate is therefore negligible compared to the mass-injection rate, providing no large over-densities occur.

The ionizing flux produced by the cluster's evolved stars, however, is both considerably more variable and higher, at between 2 $\times$ 10$^{46}$ and 8 $\times$ 10$^{47}$ ionizing photons per second. Variations are typically on timescales of the stellar death rate, as post-AGB stars go through their maximum-temperature phase over a period of a few $\times$ 10\,000 years. Shorter variations are possible as post-AGB stars go through late thermal pulses.

The variation in ionizing flux dominates the variation in the ionization parameter: the number of UV photons per injected hydrogen atom. This factor is always $\gg$1, typically $\sim$1000, meaning that intra-cluster hydrogen is permanently ionized with a very high efficiency.


\subsection{Application to other clusters}


\subsubsection{M15}

The high ionization parameter seen in 47 Tucanae should not simply be symptomatic of this cluster. For comparison, we also simulate the most-metal-poor cluster, M15. Here we take a {\sc mesa} model at 0.800 M$_\odot$ at the same $\eta = 0.41$ mass-loss efficiency and [$\alpha$/Fe] = +0.3 dex. This model reproduces a 0.672 M$_\odot$ zero-age horizontal branch star after 12.30 Gyr and a 0.555 M$_\odot$ white dwarf after 12.39 Gyr. The post-AGB star reaches a maximum luminosity of $\sim$4700 L$_\odot$ and reaches a maximum temperature of 140\,000 K. One such star was evolved along its (post-)giant branch evolution every 100\,000 years, a rate chosen to account for its fainter absolute $V$-band magnitude compared to 47 Tuc \citet{Harris10}.

\begin{figure}
\centerline{\includegraphics[height=0.47\textwidth,angle=-90]{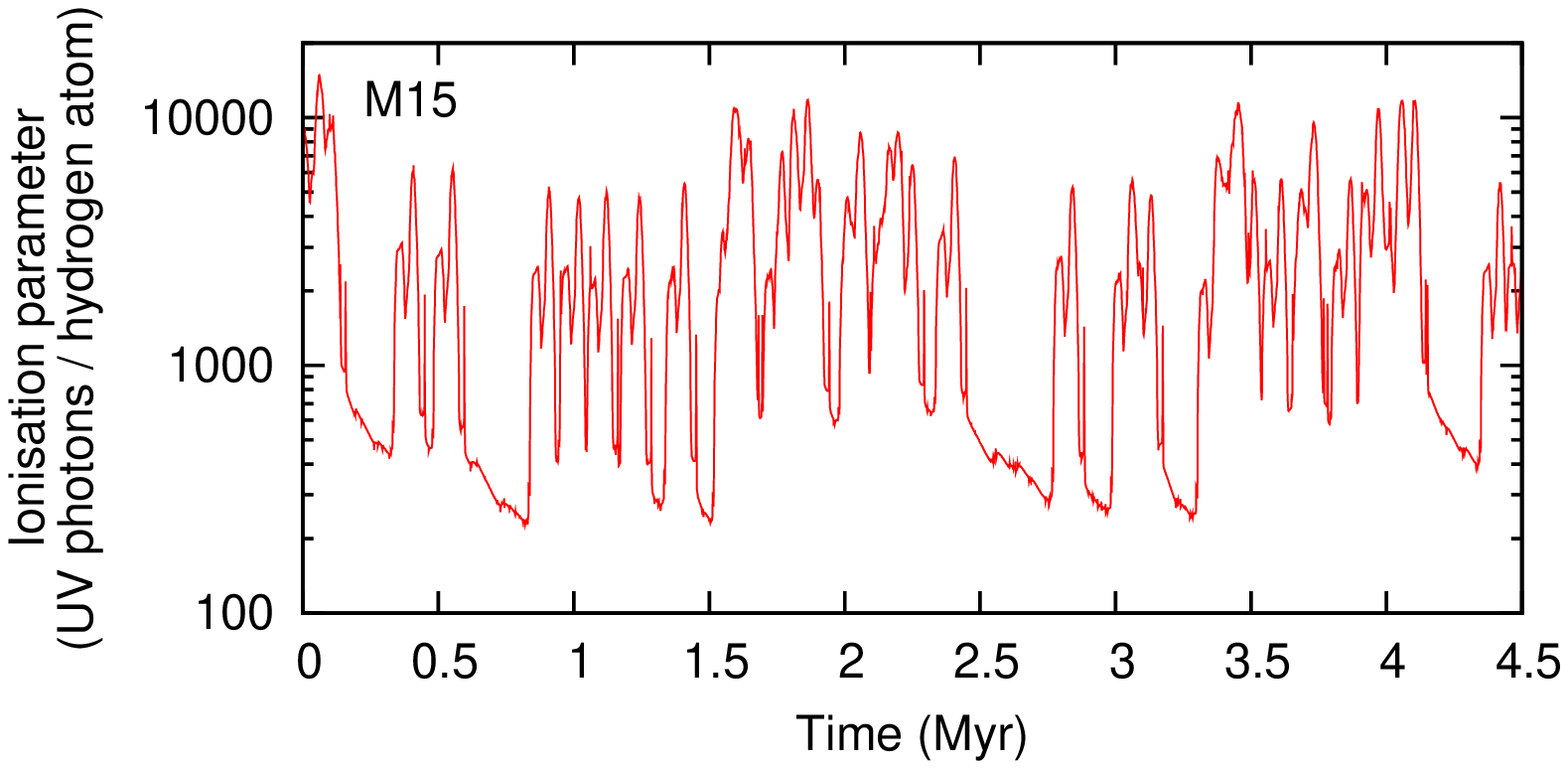}}
\centerline{\includegraphics[height=0.47\textwidth,angle=-90]{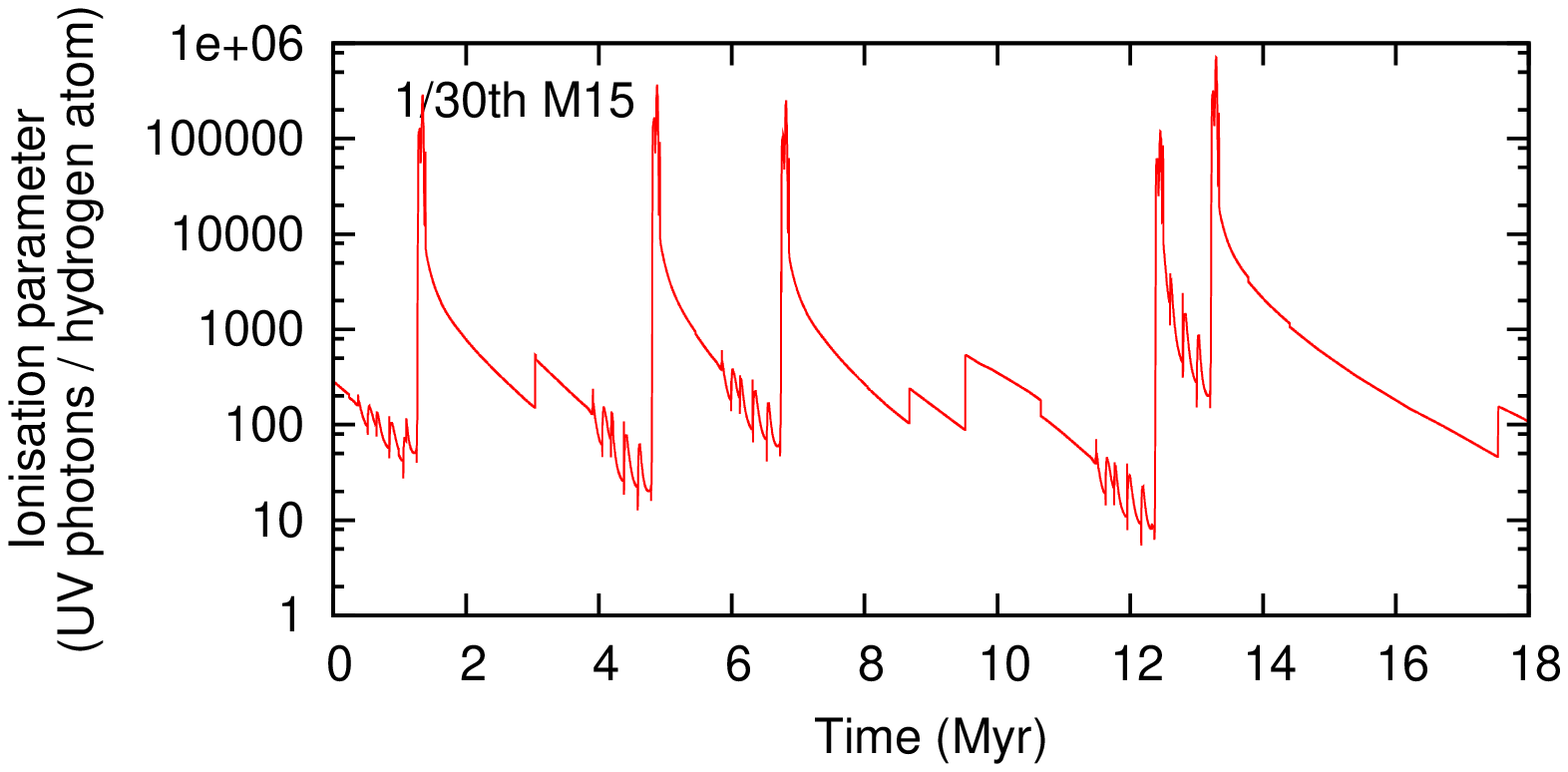}}
\caption{Top panel: expected ionization parameter in M15 as a function of time (cf.\ bottom panel of Figure \ref{IonizeFluxFig}). Bottom panel: as top panel, if M15 was 1/30th of the size (cf.\ NGC 6838). Features near 9 and 12 Myr are real and correspond respectively to a single star's RGB tip evolution and a different star's thermal pulses.}
\label{IonizeM15Fig}
\end{figure}

The resulting ionization parameter is shown in Figure \ref{IonizeM15Fig} (upper panel), and is very similar to that seen in 47 Tuc: M15 should always be ionsied. 

Yet both neutral gas and dust has been observed in M15 \citep{ESvL+03,vLSEM06,BWvL+06}. Outside of 47 Tuc, this represents the only detection of intra-cluster gas or dust to date, despite a large number of other clusters being searched \citep{MMN+08,BBW+09}. It does not appear possible for stars undergoing normal stellar evolution to inject mass into the cluster sufficiently quickly that it cannot be ionized by post-AGB stars and white dwarfs, and a sudden, recent ejection of mass may have to be invoked to explain this peculiar cloud. We return to this in Section \ref{StableSect}.

\subsubsection{Smaller clusters}

Reducing the size of the cluster has little impact on the time-averaged ionization parameter, even when the cluster size is reduced by a factor of 30. Figure \ref{IonizeM15Fig} (lower panel) shows the effect of reducing M15's mass by a factor of 30. It is unlikely that we will ever see an AGB star evolving in a cluster that is not ionized, therefore the ionization rate will always dominate over the mass-injection rate.

We note, however, that smaller clusters are characterized by long periods at relatively low ionization parameters, followed by short excursions to very high ionization parameters. In such clusters, it may be possible for localized over-densities to exist, allowing small amounts of neutral material to collect.

\section{Clearing of ICM from globulars}
\label{CloudySect}

\subsection{A pressure-supported model of the ICM}

At rest, static ionized gas should default to a gravitationally bound, pressure-supported sphere, provided the cluster clearing timescale is long compared to the free-fall timescale. To model such a system, we begin by assuming that a globular cluster can be approximated by a \citet{Plummer1911} potential. The density of the Plummer potential at a given radius, $r$, can be described as:
\begin{equation}
\rho(r) = \frac{3M}{4 \pi a^3}\left(1+ \frac{r^2}{a^2}\right)^{-5/2} ,
\end{equation}
such that the mass inside that radius is:
\begin{equation}
m(r) = M \frac{r^3}{\left(r^2+a^2\right)^{3/2}} .
\end{equation}
where $M$ is the total cluster mass and $a$ is a scaling radius.


In our assumed Plummer potential, the gravitational force on a volume element of density $\rho$ at radius $r$ becomes:
\begin{equation}
F_{\rm g}(r) = -\frac{G m(r) \rho(r) }{r^2} =  - G M \rho(r) \frac{r}{\left(r^2+a^2\right)^{3/2}}.
\label{GEq}
\end{equation}

A radiatively heated ICM must be subject to pressure support. This pressure support depends on the temperature of the plasma, and the thermalisation of that plasma. For a given plasma temperature, we can create a simple model of a pressure-supported gas system, where the force of pressure (acting outwards) is balanced by the force of gravity from the cluster's stars. From the ideal gas law ($P = nkT$), this becomes:
\begin{equation}
F_{\rm P} = -\frac{{\rm d}P}{{\rm d}r} = -kT \frac{{\rm d}n }{{\rm d}r} = -F_{\rm g} .
\label{PGEq}
\end{equation}
This equation for an element of gas can be extended to the system as a whole. Adopting the Plummer model for the gravity acting upon the gas, and replacing $\rho(r)$ by $m_{\rm H}n$ (where $m_{\rm H} \approx 1$ amu), Eq.\ (\ref{PGEq}) can be restated as the following differential equation, which must be solved iteratively:
\begin{equation}
 \frac{{\rm d}n }{{\rm d}r} =  - \frac{G M m_{\rm H} n(r) }{k T(r)} \frac{r}{\left(r^2+a^2\right)^{3/2}} ,
\label{DensityEq}
\end{equation}
creating a parameterized density gradient where only the temperature is unmodelled. However, the temperature is dictated by the cooling efficiency of the gas, which is dicatated by ionization, which is in turn dictated by both the incident radiation on the gas and the density, $n$. Further iteration is therefore required to solve the system for both $n(r)$ and $T(r)$ (see Sections \ref{InitialDensitySect} \& \ref{FinalDensitySect}).


\subsection{Application to 47 Tuc}
\label{47TucModel}

\subsubsection{A basic cluster model}
\label{Model47Tuc}

\begin{table*}
 \centering
 \begin{minipage}{160mm}
  \caption{Published and inferred properties of 47 Tuc.}
\label{ClusterTable}
  \begin{tabular}{lccl}
  \hline\hline
   \multicolumn{1}{c}{Parameter} & \multicolumn{1}{c}{Symbol} & \multicolumn{1}{c}{Value}	& \multicolumn{1}{c}{Notes and references}\\
 \hline
\multicolumn{4}{c}{Adopted published properties}\\
Total mass			& $M$		& $1.1 \pm 0.1 \times 10^6$ M$_\odot$	& \citet{LKL+10}\\
Metallicity			& [Fe/H]	& --0.72 dex				& \citet{Harris10}\\
Tidal radius			& $r_{\rm t}$	& $55.4$ pc				& \citet{Harris10}\\
Half-mass radius		& $r_{\rm 1/2}$	& $7.8 \pm 0.9$ pc			& \citet{LKL+10}\\
Plummer scale radius		& $a$		& $5.98 \pm 0.69$ pc			& \citet{LKL+10} (inferred)\\
Surrounding Halo density	& \nodata	& $\sim$0.007 cm$^{-3}$			& \citet{TC93} (very approximate)\\
Central ICM electron density	& $n_{\rm e}$	& $0.067 \pm 0.015$ cm$^{-3}$		& \citet{FKL+01} (roughly constant over $r < 2.5$ pc)\\
ICM mass in core		& \nodata	& $0.107 \pm 0.024$ M$_\odot$		& \citet{FKL+01} (integrated over $r < 2.5$ pc)\\
Stellar death rate		& \nodata	& 1 per 80\,000 years			& \citet{MBvL+11}\\
Time since Plane crossing	& \nodata	& $\sim$30 Myr				& \citet{GdJN+88}\\
Dispersion velocity		& \nodata	& $16.4$ km s$^{-1}$			& \citet{GZP+02} (at $r = 0$)\\
\ 				& \nodata	& $12.9$ km s$^{-1}$			& \citet{GZP+02} (at $r = r_{1/2}$)\\
Escape velocity			& $v_{\rm esc}$	& $68.8$ km s$^{-1}$			& \citet{GZP+02} (at $r = 0$)\\
\ 				& $v_{\rm esc}$	& $38.0$ km s$^{-1}$			& \citet{GZP+02} (at $r = r_{1/2}$)\\
 \hline
\multicolumn{4}{c}{Inferred properties from this work}\\
Mass-injection rate		& $\dot{M}_\ast$	& $2.8 \times 10^{-6}$ M$_\odot$ yr$^{-1}$	& Time average from model\\
\ 				& $\dot{N}_\ast$	& $1.06 \times 10^{44}$ amu s$^{-1}$		& Time average from model\\
UV flux 			& $\dot{N}_\gamma$	& $1-6 \times 10^{45}$ s$^{-1}$			& Time average from known sources; $\lambda < 912$ \AA\\
\ 	 			& $\dot{N}_\gamma$	& $2.43 \times 10^{47}$ s$^{-1}$		& Time average from model; $\lambda < 912$ \AA\\
Total ICM mass			& $\dot{M}_{\rm ICM}$	& $11.3$ M$_\odot$				& \\
Dynamical timescale		& \nodata		& $\sim$4.3 Myr					& Sound crossing time from $r = 0$ to $r = r_{\rm t}$\\
Thermal timescale		& \nodata		& $\sim$2270 yr					& ionization timescale of $M_{\rm ICM}$ (median value)\\
Clearing timescale		& \nodata		& $600\,000 ^{+380\,000}_{-260\,000}$ yr	& Inner 2.5 pc\\
\ 				& \nodata		& $\sim$4.0 Myr					& Entire cluster, to tidal radius\\
Outer electron temperature	& \nodata		& $\sim$12\,000 K				& Time variant, range 10\,000 -- 14\,000 K\\
Outer electron density		& \nodata		& $\sim$0.0007 cm$^{-3}$			& At $r_{\rm t}$\\
Electron column density		& \nodata		& $\sim$2.4 $\times$ 10$^{18}$ cm$^{-2}$	& \\
Total dispersion measure	& \nodata		& $\sim$0.76 pc cm$^{-3}$			& \\
ICM ejection velocity		& $v_{\rm out}$		& $\sim$10 km s$^{-1}$				& At $r_{\rm t}$, based on mass conservation\\
Sound speed			& $c_{\rm s}$		& $\sim$12 km s$^{-1}$				& At $r_{\rm t}$\\
\hline
\end{tabular}
\end{minipage}
\end{table*}

The published properties of 47 Tuc, and results obtained via this analysis, are presented in Table \ref{ClusterTable}. Three observational constraints on this model exist for 47 Tuc. The first is the central electron density from the pulsar dispersion measures, $n_{\rm e} = 0.067 \pm 0.015$ cm$^{-3}$. The second is the cluster mass $M = 1.1 \pm 0.1 \times 10^6$ M$_\odot$, and the third is the cluster half-mass radius $r_{\rm s} = 7.8 \pm 0.9$ pc, which can be converted into a Plummer scale length of $a = 5.98 \pm 0.69$ pc. A Plummer model with these parameters reproduces the observed stellar velocity distribution out to $\sim$30 pc with tolerable accuracy \citep{LKL+10}.

In applying our model of the cluster, we assume that the mass injection by stars follows the Plummer model. Mass loss from the most evolved AGB stars in 47 Tucanae is well observed. While the details of dust production by the cluster's stars are controversial \citep{ORFF+07,MBvL+11,MSS+12}, it is clear these stars produce only $\sim$1/3 of the gas which is ejected \citep{LPH+06,vLMO+06,GCB+10,MBvLZ11}. The remaining $\sim$2/3 of material is ejected by stars on the upper reaches of the red giant branch (RGB) tip \citep{GCB+10,MJZ11,Groenewegen12}. This process occurs over a longer period, hence its distribution should be more homogeneous than that of the dust-producing stars.

We note at this point that the pulsars are approximated to lie within a 2.5-pc sphere. With the above Plummer model, the fraction of the cluster's mass contained within the inner 2.5 pc is only $5.7 ^{+2.1}_{-1.4}$ per cent of the total mass of the cluster. We can therefore expect a total ICM mass considerably larger than the $\sim$0.1 M$_\odot$ previously identified \citep{FKL+01}. We also stress that our model does not include any interaction with the Halo gas: we give a fuller list of caveats to our model in Section \ref{CaveatSect}.

\subsubsection{Timescales in the central 2.5 pc}

To justify using a pressure-supported model, we must relate the clearing and freefall timescales of the inner 2.5-pc region, where boundary conditions are set by the pulsars. Using our previous assumption of one stellar death per 80\,000 years (involving a mass loss of 0.34 M$_\odot$, of which 75 per cent is hydrogen) and multiplying by the $5.7 ^{+2.1}_{-1.4}$ per cent of mass within 2.5 pc gives a mass injection rate of $1.8 ^{+0.6}_{-0.5} \times 10^{-7}$ M$_\odot$ yr$^{-1}$ of hydrogen, or $6.7 ^{+2.5}_{-1.6} \times 10^{42}$ hydrogen atoms per second, within the inner 2.5 pc. The inferred 0.107 $\pm$ 0.024 M$_\odot$ of hydrogen in the cluster core is therefore cleared on a timescale of 600\,000 $^{+380\,000}_{-260\,000}$ years.

For $a = 5.98$ pc, the inner 2.5 pc will be largely uniform ($\rho_{2.5\ {\rm pc}} = 0.49 \rho_0$), hence a typical particle will be 1.92 pc from the centre and initially experience an inward acceleration of $5.9^{+3.1}_{-2.0} \times 10^{-10}$ m s$^{-2}$. Integrating over time, a free-falling body would achieve a velocity of 8.7 km s$^{-1}$ upon arrival at the cluster centre after 350\,000 years. Since the 600\,000-year clearing timescale is longer than the 350\,000-year freefall timescale for material in the inner regions, a pressure-supported ICM appears justified, even before we consider internal heating.

\subsubsection{Timescales of the entire cluster}

In order to correctly model the cluster, we need to further consider which parameters we can consider static, and which will be time variable. If the ICM can respond to changes to a parameter on timescales shorter than $\sim$10\,000 years (the timescale over UV flux change from a new post-AGB star), then that response can effectively be considered instantaneous and that parameter can be considered time variable. If the ICM cannot respond to changes within $\sim$300\,000 years (the time over which UV flux averages out to a constant) then a static value is appropriate for that parameter. Responses on intermediate timescales require more-detailed modelling, but there are no parameters for which this is important.

Recombination timescales in the central region can be expected to be several Myr, increasing in time as one moves out to lower-density environments. The total mass of the ICM implied in the cluster is a few solar masses ($\sim$10$^{57}$--10$^{58}$ atoms) and the typical mass injection rate is $\approx$1 $\times$ 10$^{44}$ atoms s$^{-1}$, implying a cluster clearing timescale of $\sim$1 Myr (we revise this in the following section), meaning the density structure of the ICM will be largely static. However, the typical ionizing flux is $\sim$10$^{47}$--10$^{48}$ photons s$^{-1}$, suggesting a thermal equilibrium timescale of $\sim$100--1000 years (we revise this later). We can therefore expect that the temperature structure of the ICM can respond to the changes in UV flux, even if the density structure cannot. We caution, though, that while we expect the density structure of the entire cluster's ICM to be static, changes in both UV flux and mass-loss rate in small (sub-pc-size) regions may occur dynamically.

\subsubsection{An initial {\sc Cloudy} model of the ICM density}
\label{InitialDensitySect}

\begin{figure}
\centerline{\includegraphics[height=0.47\textwidth,angle=-90]{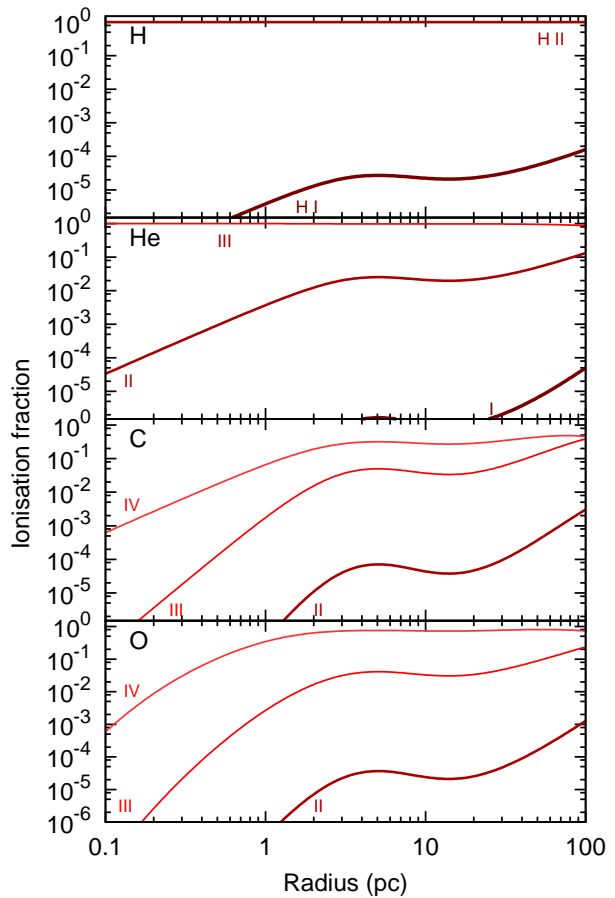}}
\caption{ionization states, as modelled by {\sc Cloudy}, for the time-averaged ionizing flux predicted for 47 Tuc (Section \ref{MESASect}). The labels refer to the ionization state of the corresponding ion: lines become thinner and lighter for higher ionizations. Dominant species are: H {\sc ii}, He {\sc iii}, C {\sc iv} or higher, and O {\sc iv}, throughout the entire cluster.}
\label{CloudyIonFig}
\end{figure}

\begin{figure}
\centerline{\includegraphics[height=0.47\textwidth,angle=-90]{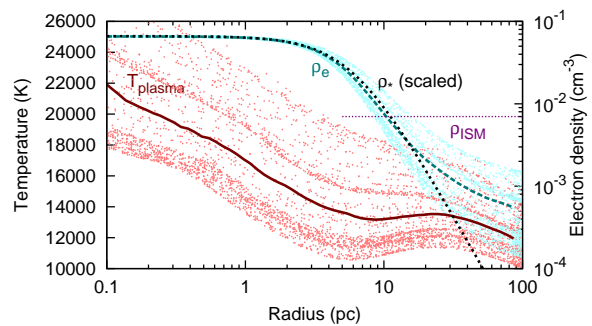}}
\caption{Temperature and density profiles for ICM in 47 Tuc as modelled by {\sc Cloudy}, from {\sc MESA} evolution and {\sc bt-settl} atmosphere models. Realisations for 220 ionizing fluxes are shown, representative of different circumstances in the evolution of cluster's post-AGB population. The dashed/cyan $\rho_{\rm e}$ line shows the final density profile that was adopted for the cluster's ICM. The adopted Plummer distribution for stars and anticipated Halo ISM density are also shown.}
\label{CloudyFig}
\end{figure}

The final missing parameter from Eq.\ (\ref{DensityEq}) is the plasma temperature. The {\sc Cloudy} code allows one to model the photo-ionization of ICM for a given density distribution, under certain assumptions. The most significant of these assumptions is that the environment is static: i.e., that there is no outflow\footnote{A wind solution is programmed into {\sc Cloudy}, but the case of a positive (outflow) velocity has not been fully tested.}, and that material is ionized {\it in situ}. {\sc Cloudy} also does not fully model kinetic energy input into the gas, hence we assume that particle winds (e.g.\ from main-sequence stars, less-evolved red giants, horizontal branch stars and pulsars) do not heat the ICM. Since the effect of both of these limitations is to reduce the energy available and its transport to large radii, they provide a conservative case where all the available energy for ejection is derived from radiation from the white dwarfs.

One outstanding issue is the treatment of cooling by expansion and gravity, which is not accounted for in {\sc cloudy}. Gravitational cooling may arise as particles flow outward from the cluster, while expansion cooling occurs when material expands as it flows to larger radii where densities are lower. However, if we simply wish to show that mass can escape the cluster, this reduces to the case that the outward velocity is small compared to the thermalisation timescale, thus it can be neglected. Since the thermalisation timescale is $\sim$100--1000 years, a modest ($\sim$10 km s$^{-1}$) wind will not experience an appreciable change in gravitational potential energy during this time (typically $\sim$0.01--0.1 eV, compared with the $\sim$13.6 eV of incoming radiation).

Our only fixed boundary conditions are the time-varying ionizing flux and central density (assumed constant). Solving Eq.\ (\ref{DensityEq}) must therefore be done outwards and iteratively, using some form of time-averaging of the ionizing radiation to produce a static density structure. We begin with a model of the ionizing flux equal to the time-average of the radiation field from highly evolved stars: a total flux of $2.43 \times 10^{47}$ photons s$^{-1}$, with a characteristic temperature of 65\,000 K (Section \ref{MESASect}; Figure \ref{IonizeFluxFig}). For the {\sc Cloudy} input file, this was broken down into the time-averaged flux coming from stars in each 1000-K bin. We also begin with a spherical geometry and a flat hydrogen density of $\log n_{\rm H} = -1.17$ (cm$^{-3}$) across the entire cluster out to 100 pc. We assume the ICM has the same abundances as {\sc cloudy} prescribes for a standard H {\sc ii} region, depleting the metals as appropriate for 47 Tuc ([Fe/H] = --0.72) and depleting dust-forming elements into grains. 


With this model, we use {\sc Cloudy} to compute the temperature profile for the cluster. We then solve the density profile using Eq.\ (\ref{DensityEq}) and recalculate the temperature profile, iterating until we converge to a solution.

\subsubsection{A final static density distribution}
\label{FinalDensitySect}

The ionization of various species in this model are shown in Figure \ref{CloudyIonFig}. From these figures, it is clear that the ionization can be maintained to a high degree among several species, out to radii exceeding the tidal radius of the cluster (55 pc; \citealt{Harris10}).

We have so far assumed a simple time-average of the radiation to obtain a density profile for ICM. However, this time average is never uniquely achieved, due to the distribution of characteristic temperatures and ionizing fluxes present over time in the cluster (Figure \ref{IonizeFluxFig}, bottom panel): the characteristic temperature and ionizing flux never approach the average values simultaneously. We can use our initial model as a starting point to compute a more accurate density distribution.

To do this, we have taken 220 different timesteps of our model, each with its own set of ionizing sources. For each timestep, we make a new {\sc Cloudy} input file using these as the heating sources and taking our previously-calculated density distribution as a starting point. We then rerun {\sc Cloudy} to create a temperature model for that input file, re-solve the density profile, and iterate as before to establish a density profile which would be appropriate for a static system with the input ionizing sources and observed central electron density. An average density model (Figure \ref{CloudyFig}) was then computed as an average of all 220 profiles, and we take this as our final static density distribution.

Perhaps unsurpringly, the gas density closely follows the stellar density. Only after $\sim$20 pc does the gas density becomes divergently larger due to the lower gravity. A density of $\sim$0.0007 amu cm$^{-3}$ is found at the tidal radius. A total column density of $\sim$2.4 $\times$ 10$^{18}$ electrons cm$^{-2}$ is found, giving a total dispersion measure of 0.76 pc cm$^{-3}$ through the cluster. Notably, a total of 11.3 M$_\odot$ of ionized ICM is inferred within the tidal radius, as much as produced by the stars of 47 Tuc over 4 Myr (the cluster last crossed the Galactic Plane $\sim$30 Myr ago; \citealt{GdJN+88}).

This is much greater than the ICM mass and clearing timescales inferred for other clusters ($\lesssim$1 Myr; e.g.\ \citealt{BMvL+08}), primarily because these studies have considered only neutral ICM. Ionized ICM, by comparison, is largely invisible. It is also larger than predicted by other mechanisms. \citet{Smith99} predicts a much faster outflow from a higher-temperature wind (59\,300 K). This predicts (his eq.\ (12)) that the central density is $\sim$0.001 amu cm$^{-3}$, rather than the $\sim$0.067 amu cm$^{-3}$ observed. This discrepency may have come from the assumption in \citet{Smith99} that the cluster's main-sequence stars have winds with the same velocity and mass-loss rate as the Sun, whereas at least the mass-loss rate is likely to be lower for these older stars (e.g.\ \citealt{WMZ+05}). Conversely, it is much lower than the $\sim$1--6 per cent of the cluster's mass (i.e.\ $\sim$10$^4$ M$_\odot$) that is needed before it can trigger further star formation \citep{NRRL12}.

The Kelvin--Helmholtz timescale for the system can be calculated as the total thermal energy in the ICM divided by the ionizing flux. We calculate this as $\sim$330 years, sufficiently smaller than the timescale for change in the ionizing radiation that the ICM should respond to thermal changes. The dynamical timescale can be calculated as the sound-travel time between the cluster core and outer edge of the cluster. For an outer edge equal to the tidal radius, this is $\sim$4.3 Myr. Given the accuracy of the observations, this is identical to the clearing timescale. It is also sufficiently greater than the timescale over which ionizing radiation can be considered constant that a time-invariant ICM density seems appropriate.

We remind the reader that we do not consider the complex interaction of 47 Tuc with the Halo gas. We note that our model reaches the excepted density of the surrounding Halo gas ($\sim$0.007 cm$^{-3}$; \citealt{TC93}) after only $\sim$10 pc. In reality, this density is poorly known, and the bow shock stand-off radius depends on equating the pressure of the incoming Halo gas with the combined thermal and bulk-flow pressure of the outgoing ICM (see, e.g., \citealt{CKvM+12}, and references therein). We cannot adequately model this within a static simulation, hence for the remainder of this paper we assume the bow shock to lie at or outside the tidal radius.


\subsection{Tidal escape from the cluster}

\subsubsection{Application to 47 Tuc}

\begin{figure}
\centerline{\includegraphics[height=0.47\textwidth,angle=-90]{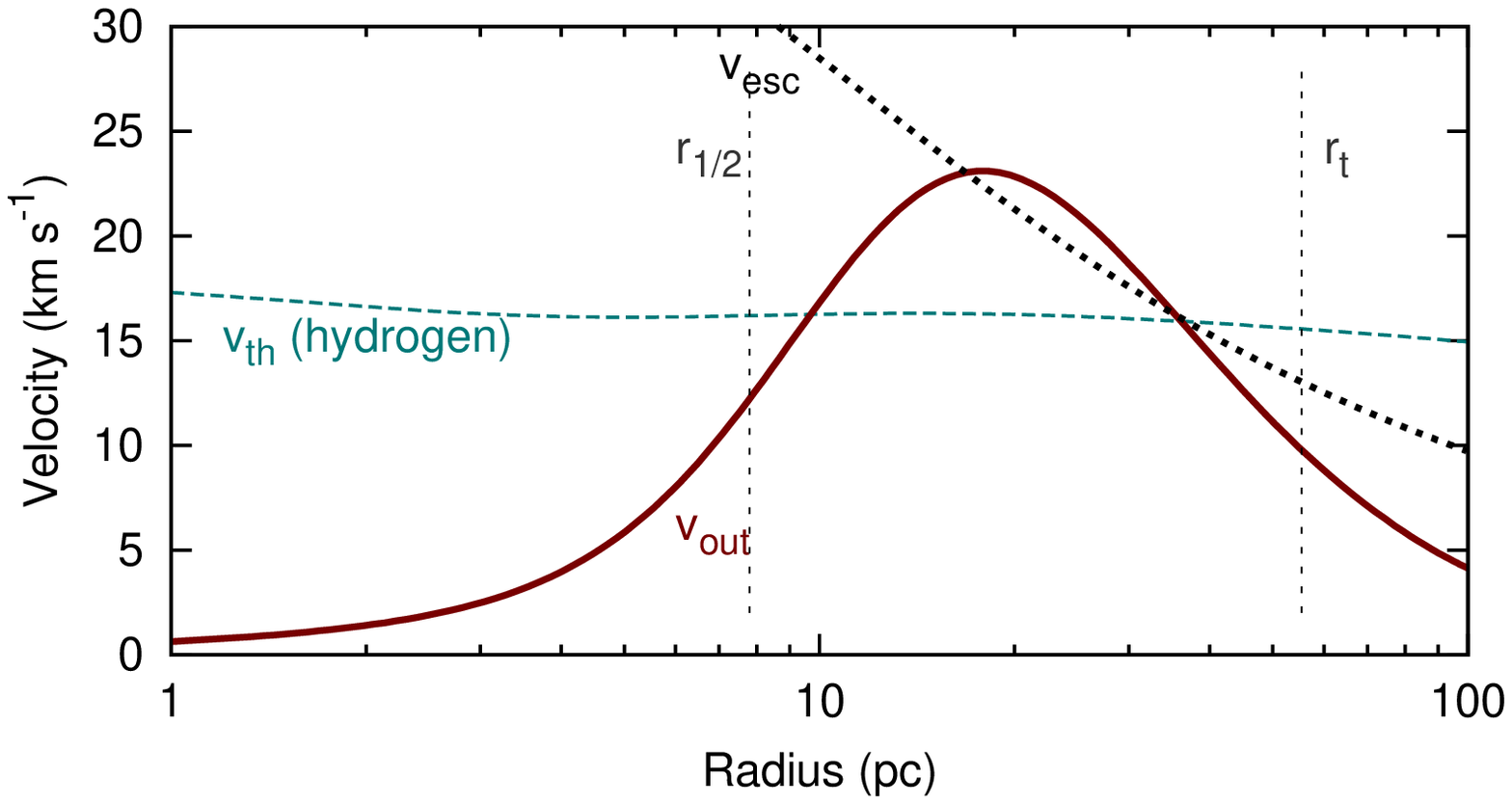}}
\centerline{\includegraphics[height=0.47\textwidth,angle=-90]{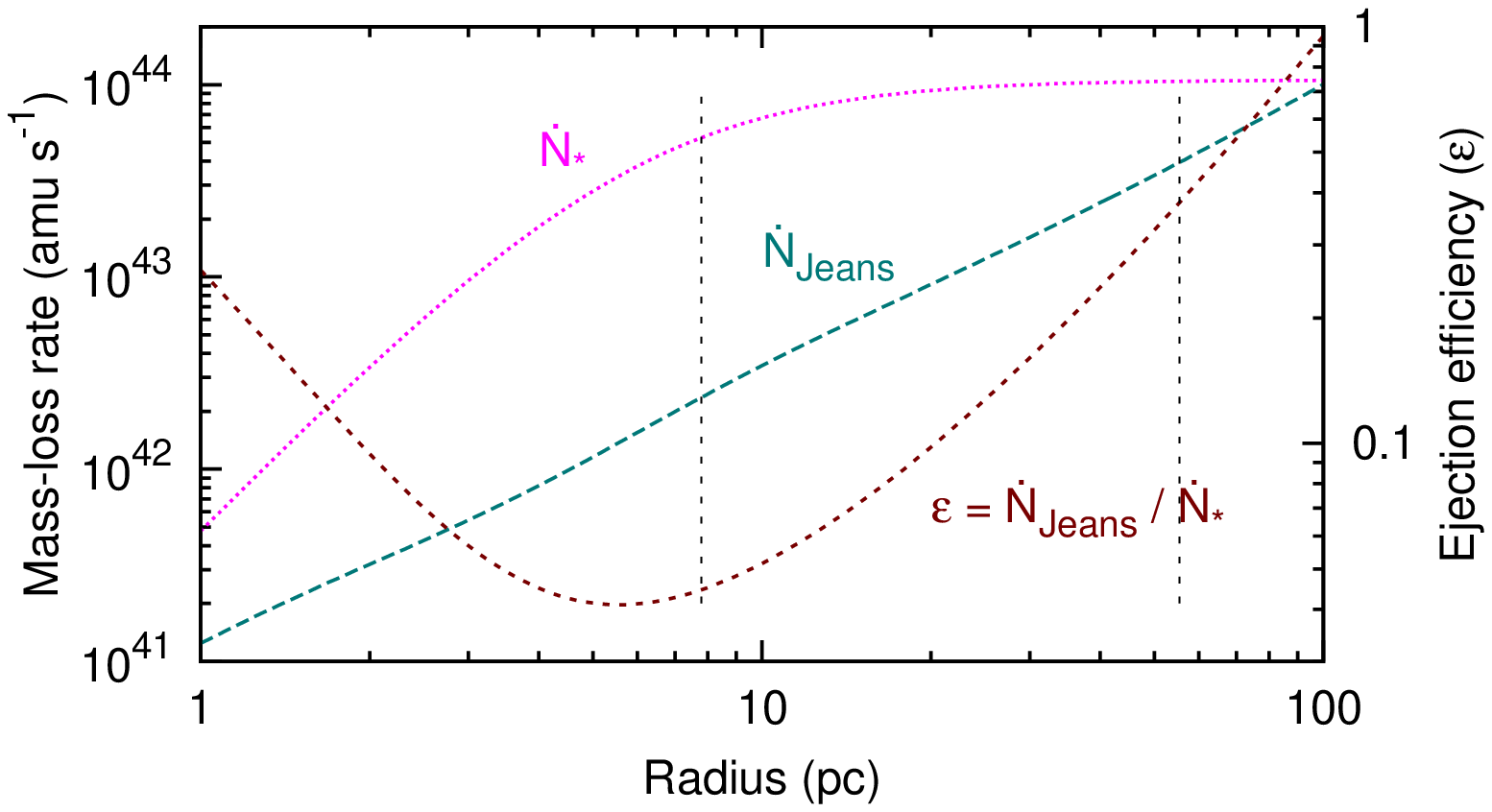}}
\caption{Top panel: Variation of different velocities with radius. The mean thermal velocity of hydrogen  ($v_{\rm th}$) and escape velocity ($v_{\rm esc}$) are shown, along with the wind velocity ($v_{\rm out}$) required to maintain a constant electron density profile (Figure \ref{CloudyFig}) given that mass is injected into the ICM at a constant rate of 1.06 $\times$ 10$^{44}$ atoms s$^{-1}$. Bottom panel: particle-injection rate ($\dot{N}_\ast$) and particle-loss rate ($\dot{N}_{\rm Jeans}$) as a function of radius, were the gas able to undergo Jeans escape at that radius. The vertical dotted lines represent the half-mass and tidal radii.}
\label{CloudyvFig}
\end{figure}

\begin{figure}
\centerline{\includegraphics[height=0.47\textwidth,angle=-90]{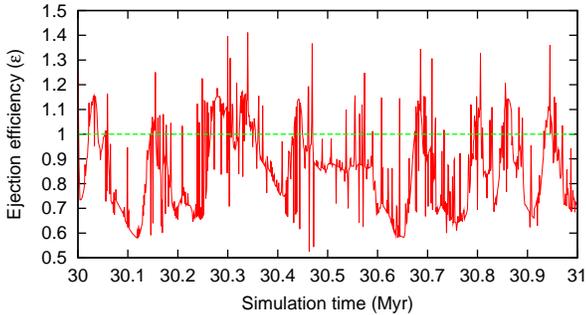}}
\caption{Ejection efficiency ($\dot{N}_{\rm Jeans}/\dot{N}_\ast$) at the tidal radius for a constant density cluster model of 47 Tuc with a time-varying ionizing flux: $\epsilon > 1$ denotes ICM is lost from the cluster.}
\label{CloudyepsFig}
\end{figure}

We discussed in Section \ref{PhysEscSect} that material should first escape the cluster tidally, then by Jeans escape through the truncated outer boundary of the cluster. Solving Eq.\ (\ref{TidalEq}) for the calculated density ($\sim$0.0007 cm$^{-3}$) and sound speed ($\approx$12 km s$^{-1}$ for $T \approx 13\,000$ K; Figure \ref{CloudyFig}) at the tidal radius of 47 Tuc (Table \ref{ClusterTable}) shows that only very small ($\sim$8 AU) extensions beyond the tidal radius can occur before $\dot{M}_{\rm tidal}$ exceeds the mass-injection rate by stars ($\sim$2.8 $\times$ 10$^{-6}$ M$_\odot$ yr$^{-1}$). Hence we expect the ICM to be strongly truncated at the tidal radius, and tidal outflow to be very effective. However, we must also consider the ability of material to escape via Jeans escape.

Figure \ref{CloudyvFig} shows the thermal and escape velocities derived from our time-averaged model. The escape velocity at the cluster centre (68 km s$^{-1}$; \citealt{GZP+02}\footnote{Online table at:\\ http://dept.astro.lsa.umich.edu/$\sim$ognedin/gc/vesc.dat}) is not well reproduced by the Plummer model, which gives 40 km s$^{-1}$, however escape velocities beyond the half-mass radius are relatively well reproduced. The thermal velocity is relatively constant, being maintained at $\approx$15 km s$^{-1}$ over all radii $\gtrsim$1 pc, thus the Jeans escape parameter is always $\lambda_{esc} \lesssim 6$ beyond the half-mass radius, and $\lesssim 15$ in the cluster centre. This can be compared with the $\lambda_{esc} \lesssim 6$ typically assumed for Jeans escape of planetary atmospheres (e.g.\ \citealt{CZ09}). The criterion $\lambda_{esc} \approx 1$ is reached inside the tidal radius, implying that Jeans escape becomes hydrodynamic once the density drops at the tidal radius, and the ICM boils off the cluster.

Under the assumption of a time-invariant density profile, one can use Eq.\ (\ref{JeansvEq}) to calculate the outflow velocity. This velocity, $v_{\rm out}$, is also plotted in Figure \ref{CloudyvFig}. We find an outflow velocity of $\sim$10 km s$^{-1}$ must be sustained from the cluster. The velocity of this wind becomes mildly supersonic over much of the cluster, and exceeds the escape velocity between $\sim$13 and 40 pc. At the tidal radius, $v_{\rm out} \approx 10$ km s$^{-1}$ (Figure \ref{CloudyvFig}), implying material that is tidally lost will have a relatively slow bulk ejection velocity. While this is a substantial fraction of the escape velocity, it implies that the wind emanating from the cluster may move quite slowly, which has implications for its interaction with the Halo gas.

The bottom panel of Figure \ref{CloudyvFig} shows how $\dot{N}_{\rm Jeans}$ (Eq.\ (\ref{JeansEq})) and $\dot{N}_\ast$ (Section \ref{MinIonRate}) evolve as a function of radius for our static, time-averaged model. The quotient gives the efficiency with which material is ejected from the system, which we define as:
\begin{equation}
\epsilon = \frac{\dot{N}_{\rm Jeans}}{\dot{N}_\ast} ,
\end{equation}
and which is also plotted in Figure \ref{CloudyvFig}. Efficient ejection $\epsilon = 1$ is only achieved at a radius of $\sim$100 pc, and it remains a factor of $\sim$3 short at the tidal radius. However, we recall our earlier statement: while the density distribution of the model is inflexible to changes in the ionization rate, the temperature structure of the model is time-variant. Hence, we now extend our model into the time domain.

To do this, we have taken our fixed-density {\sc cloudy} model, and changed the ionizing radiation field to reflect different times in our simulated cluster. We calculate 1000 models, covering the cluster's evolution over 1 Myr in 1000-year time steps. For each time step, we calculate the $\dot{N}_{\rm Jeans}$, $\dot{N}_\ast$ and ejection efficiency, $\epsilon$, at the tidal radius. The efficiency is plotted in Figure \ref{CloudyepsFig}. In our model, ejection efficiencies exceed unity for 22 per cent of the time. During these periods, heating of ICM by white dwarfs is sufficient to eject material from our modelled cluster simply by driving the thermal equilibrium of the ICM.

The average efficiency in our model is 86 per cent. Given the uncertainties in the ICM central density, the final stages of evolution of the cluster's white dwarfs and the precise injection rate of material by stars, this is tolerably close to 100 per cent. With the Jeans mass-loss rate being so high, we therefore expect the ICM to escape at a much higher efficiency than it is tidally lost through the $L_1$ and $L_2$ points. The closeness of the efficiency to unity is sufficient to state that, without any additional kinematic or heating mechanisms, heating of ICM by white dwarfs is sufficient to cause catastrophic ejection of ICM from the cluster in our model. Our model therefore provides both a mechanism for clearing globular clusters of ICM, and an estimate of the ICM mass in 47 Tuc (Section \ref{FinalDensitySect}) which matches that inferred from observations.

\subsubsection{Unmodelled heating, cooling and removal mechanisms}
\label{CaveatSect}

In this Section, we have presented a scenario whereby the ICM is heated by white dwarfs to a sufficient extent that the ICM cloud can expand across the tidal radius, after which material is lost by a combination of Jeans escape and tidal escape, where the former dominates. We caution that we do not provide a full hydrodynamic model of the cluster. Several factors which may be physically important in governing the cluster's ICM are not included in our simple model and its assumptions. We detail those we consider most important here.

{\it Hydrostatic equilibrium and no gravitational or expansion cooling:} Our {\sc cloudy} model assumes hydrostatic equilibrium. The temperature, density and ionization profiles are therefore calculated under the assumption that the ICM is ionized in situ. The recombination timescale is $\sim$Myr in the inner regions of the cluster and increases as density drops. Gravity is only included in our model as a pressure-balancing term and cooling by expansion of gas is not acounted for. Outflowing material will expand as it moves outwards and lose energy as it overcomes the gravitational potential and expands. While we consider this not to be important (Section \ref{InitialDensitySect}), it is unmodelled. Conversely, acoustic transport of energy outwards may also occur, causing heating of material further out, but this should serve to increase the cluster's mass-loss rate.

{\it Time-varying tidal radius:} We have assumed that the ICM mass can be calculated from the dynamical timescale of the tidal radius, which is likely to be $\sim$10$^6$ Myr for a typical globular cluster. Near the Galactic Plane, the tidal radius of a cluster can change appreciably on these timescales, though this is unlikely to affect the particular case of 47 Tuc.

{\it A point ionization source:} We model our input ionization as a single, point source lying in the cluster centre. In reality, ionizing sources should be distributed roughly according to the Plummer mass distribution, with some degree of central concentration due to mass segregation. Known sources have a strong central condensation (Figure \ref{MapFig}), but this is partly due to survey biases. While this has implications for the ionization conditions in the innermost parts of the cluster, the tidal radius is sufficiently far from the cluster centre that ionizing sources can be treated as a point object.

{\it Parametric uncertainties:} Our model adopts the best-estimate values on the Plummer scale radius, cluster mass, central electron density, and other, less important quantities. We have also fixed a Plummer model as the density distribution.

{\it No diffusion:} We do not consider diffusion of different ionic species within the gas, nor any fractionation of elements this produces, nor any outward thermal energy transport that occurs. Such a treatment requires more detailed study of magnetic properties of the plasma.

{\it Stellar velocity dispersion:} We have assumed that mass is injected into the ICM by a static, homogeneous population of stars following the Plummer potential. In reality, stars have an additional dispersive velocity: $\approx$16 km s$^{-1}$ at the cluster centre, decreasing to $\approx$13 km s$^{-1}$ at the half-mass radius \citep{GZP+02}. An additional velocity component comes from the expansion velocity of the stellar winds (thought to be $\sim$10 km s$^{-1}$ for most evolved globular cluster giants; e.g.\ \citealt{MvL07}). Material ejected from these stars has more momentum than modelled. This momentum will disperse in turbulent motion in a static ICM cloud, leading to an additional heating term which will cause the expansion of the cloud, increasing the mass-loss rate from its outer edge.

{\it Stellar evolution variation:} We have not accounted for the fact that a variety of evolutionary outcomes are possible for stars, even in the simple stellar population of a globular cluster. A small percentage of binary stars will merge, becoming blue stragglers. These will have a higher luminosity and will be able to ionize more ICM. A variety of helium abundances may be present in globular clusters. Helium-rich stars evolve faster, thus helium-rich AGB stars are presently of lower mass. These produce lower-luminosity post-AGB stars with less ionizing radiation. Of all the massive globular clusters, 47 Tuc has one of the most homogeneous populations, so this should not be a large effect \citep{GCB+10,CPJ+14}.

{\it Other heating mechanisms:} A variety of additional heating mechanisms are possible, which will have a similar effect. These include all the previously considered heating sources for ICM in globular clusters: blue horizontal branch stars (though these are not so relevant for 47 Tuc), main sequence stellar winds, pulsar winds, etc.

{\it Halo gas shocking:} We have yet to consider that the material can also be removed from the cluster by bow shocking by interstellar gas. \citet{PRS11} show that the bow shock around a 47-Tuc-like cluster is expected to approach within a few parsecs from the cluster core. However, this was for an internal temperature of $<$10\,000 K and a density for Halo gas which is larger than the value typically assumed in the region of 47 Tuc ($\sim$0.007 cm$^{-3}$; \citealt{TC93}) by a factor of around 100. The combination of these factors would suggest that the true stand-off radius is likely to be considerably further from the cluster core. We suggest that another look at the hydrodynamic models would be helpful in understanding the interplay between the intracluster and Halo gas.


\section{Testing the limits of an ionized ICM}
\label{StableSect}

We have re-run our final {\sc cloudy} model, changing the cluster parameters in turn, and again iterating until we reach a stable solution. In some cases, a solution could not be reached when the change in temperature with the tested parameter became too great. Examination of these cases showed that this happens when recombination becomes important. At this point, the ICM rapidly collapses into a neutral state, where it has the potential to build up in within cluster, possible forming another generation of stars.

Without a prediction of a central ICM density or total ICM mass, we cannot make more than broad estimates, based on taking our model for 47 Tuc and scaling certain parameters. We therefore caution the reader that these results are not exact predictions of clusters with parameters changed in this way.

\subsection{Tidal radius}

Figure \ref{CloudyvFig} shows that the implied Jeans-escape efficiency ($\epsilon$) is fairly small at small radii within 47 Tuc, decreases to a minimum at around $\sim$7 pc, then increases to exceed equality at $\sim$100 pc. For a fixed cluster mass, we can naively predict that tidal escape through the Lagrangian points may be more important for clusters with smaller tidal radii, however this is modified by the ICM density (Eq.\ (\ref{DensityEq})). The ICM sound speed is largely set by the (largely fixed) temperature of the ionizing sources, the total ICM mass should be related linearly to the tidal radius. However, the enclosed volume will be $\propto r_{\rm t}^3$, hence the average ICM density will vary as $\rho \propto r_{\rm t}^{-2}$. Perhaps unsurprisingly, larger, more-massive clusters can be expected to have denser ICM.

\subsection{Cluster density}

To test sensitivity to cluster density, we began with our time-averaged hydrostatic model. We constructed a series of models with an identical total cluster mass and ICM mass, but different Plummer radii. We allowed the system to reach a stable density and temperature structure. Diffuse clusters had correspondingly diffuse ICM, with a higher density, higher mass-loss rate but only marginally lower temperature at the tidal radius (kept constant at 55.4 pc). Models for densities a few times that of 47 Tuc did not converge, as they became divergent when recombination occurred. We therefore estimate that clusters a few times denser than 47 Tuc (but of the same mass) should retain their ICM.

Though we do not model them here, changes to the shape of the cluster potential will also impact the retention of ICM. A key factor is the central density of the cluster, as this sets the recombination efficiency. Increasing the stellar density corresponds to an increase in central escape velocity as $v_{\rm esc} \propto \sqrt{\rho}$. The difference between our modelled Plummer potential ($v_{\rm esc} = 40$ km s$^{-1}$) and that calculated by \citep[][; 68.8 km s$^{-1}$]{GZP+02} corresponds to a density increase by a factor of three. The velocities at the half-mass radius match more closely (38 and 32 km s$^{-1}$, respectively). Taking the \citet{GZP+02} potential, it would be possible that clusters of similar mass which are only slightly more dense than 47 Tuc could retain their ICM.

\subsection{Cluster mass}

In this test, we kept a constant cluster density and tidal radius, and scaled the total and ICM masses. We again iterated the density and {\sc cloudy} models to reach a stable density and temperature structure.

Clusters less massive than 47 Tuc had a lower central density, but higher density and temperature beyond $\sim$40 pc, leading to faster mass loss. A model with twice the mass of 47 Tuc shows $\sim$1 per cent neutral hydrogen in the centre, with a lower total mass-loss rate than 47 Tuc. Clusters more than three times the mass of 47 Tuc did not converge as recombination became important. We therefore estimate that clusters more than $\sim$3$\times$ the mass of 47 Tuc would retain their ICM.

\subsection{Cluster age or stellar mass}

Additional {\sc mesa} models were calculated for stellar masses of 1.15 and 2.00 M$_\odot$ and synthetic 47-Tuc-like clusters calculated. In higher-mass stars, the post-AGB evolution is faster and the mass-loss rate higher. Hence, the average ionization parameter (number of UV photons per hydrogen atom) decreases as stellar mass is increased. From our present-day, 0.899 M$_\odot$ model has an average ionization parameter of 2300. This dropped to 800 for the 5 Gyr, 1.15 M$_\odot$ model and 300 for the 1 Gyr, 2 M$_\odot$ model. In the 2-M$_\odot$ model, the ionization parameter at times fell to as low as 20. Depending on the central ICM density, stellar density and tidal evaporation of the cluster during the last 10 Gyr, we predict that the ICM of 47 Tuc may have reached neutrality around or shortly before 1 Gyr in age. We cannot accurately define this age at present: ionization by main sequence stars is not considered here, but is expected to decline with cluster age, reaching the UV flux in the solar neighbourhood after $\sim$250 Myr (4 M$_\odot$; Zhukovska et al., in prep.). Changes in the stellar death rate due to stellar evolution and cluster evaporation have also not been considered.

\subsection{Application to other clusters}

\subsubsection{Present clusters}

Using these relations, we can qualitatively predict what happens when our model is perturbed. In our model, the ICM remains permanently ionized and simply expands dynamically to fill the void left by material boiled off its tidal surface. We can expect this process to happen provided a sufficient density can be maintained at the tidal radius for either Jeans or tidal escape to be effective. This is expected to be true for any ionized system: if material is not lost, then density will increase, recombination will increase, and extra ionizing photons will be absorbed, heating the system, increasing pressure support and expanding the cloud.

However, if ICM density is allowed to increase sufficiently that all the ionizing photons are absorbed, the Str\"omgren sphere will contract to within the cluster and the ICM will become neutral. The temperature of the ICM will decrease, the escape rate of material will decrease, ICM density will increase, boosting the recombination rate. This results in a situation whereby material can gravitationally sink to the centre of the globular cluster and build up a larger neutral ICM.

Few (if any) present-day clusters should have the required combination of mass and density to maintain and retain a neutral ICM, even in small, localized regions. The strongest candidates are massive, core-collapsed clusters, of which M15 is among the best candidates. In the present day, this would require a catastrophic increase in localized mass-loss rate from stars, e.g.\ through the instantaneous ejection of a substantial part of a star's atmosphere during a stellar encounter or binary merger. A localized over-density in the ICM could allow a cloud of ICM to persist over a short period of time. We put this forward as an explanation for the small ICM cloud in M15 \citep{ESvL+03,BWvL+06}. We therefore predict that this should be quickly evaporating.

\subsubsection{Past clusters}

In the past, increasing ICM density to the point where the ICM can be retained should have been easier. No globular clusters currently exist with masses more than the critical $\sim$3$\times$ the mass of 47 Tuc. \citet{GZP+02} find NGC 6388 to be 1.45$\times$ as massive as 47 Tuc, M54 to be 1.73$\times$ as massive and $\omega$ Cen to be 2.2$\times$ as massive. M54 is uniquely placed in the heart of the Sgr dSph galaxy, thus has a peculiar history (e.g.\ \citep{SDM+07}). Both of the other clusters, NGC 6388 and $\omega$ Cen, have some of the best examples among Galactic globular clusters of multiple populations (e.g.\ \citealt{Piotto09}). Typically, these are manifest as stars with differing light element abundances (C, N, O, ..., Si). However $\omega$ Cen also shows a spread in [Fe/H], and is sometimes considered a `transition object': a dwarf galaxy nucleus that has shed its dark matter and stellar halo either through internal mass loss or external tidal disruption \citep{LJS+99,BF03}. Such objects blur the line between globular clusters and galaxies, which may be best defined by the presence of dark matter \citep{WS12}.

Globular cluster systems originally had several times more mass than their modern-day counterparts. This was subsequently lost as stars were tidally stripped or underwent mass loss in the intervening $\sim$10--13 Gyr. Ejecta from massive stars, which underwent Type II supernovae, had sufficient velocity to escape the proto-cluster, hence did not typically increase [Fe/H] beyond the cluster's nascent value. Current models explain the differing pattern of light elements pattern by preferentially retaining material from heavier ($\sim$3--8 M$_\odot$) AGB stars, which lose mass in the few $\times$10$^{8}$ years after the cluster's formation \citep{CS11b,DEVDA+08,DEDAV+10,DEDAC+12}. This ejecta can then collect in the centre of the young clusters, forming subsequent generations of stars, interrupted by occasional Type Ia supernovae, whose ejecta kinematically cleared the clusters (including $\omega$ Cen; \citealt{JP10}).

Our above calculations predict that clusters only need to be a few times more massive, or have stars a few times the mass of the Sun, in order for the ICM to be retained by the cluster. The more-massive, present-day Galactic globular clusters would therefore appear to have met both of these criteria in their early existences. We suggest these systems were able to retain their ICM, allowing the formation of more than one generation of stars. More-massive clusters should retain their ICM for longer, producing several generations and yielding a larger light-element abundance spread. These clusters will become `transition objects' (cf.\ \citealt{WS12}): objects with broad spreads in [$Z$/H] (or even [Fe/H]) but without necessarily having the dark matter halos of dwarf galaxies. We therefore present a new mass hierarchy, supplementing that of \citet{WS12} to include clusters' enrichment history:
\begin{enumerate}
\item the lowest-mass globular clusters should never have had sufficient mass to retain their ICM, forming a single generation of stars; 
\item higher-mass globular clusters should have had a sufficiently deep gravitational potential to retain the winds of intermediate-mass AGB stars ($\sim$3--8 M$_\odot$), forming multiple-population clusters; 
\item the highest-mass globular clusters, principly $\omega$ Cen, were massive enough to retain Type II supernovae ejecta, but not massive enough to attain/retain dark matter halos, forming `transition objects'; 
\item objects more massive than this were also able to retain dark matter halos and probably also Type Ia supernovae ejecta, forming dwarf galaxies.
\end{enumerate}

Comparisons of stellar yield calculations with the spread in light and heavy elemental abundances of stars in clusters can tell us the cut-off point beyond which ICM was ionized and able to escape the cluster. This would help refine this hierarchy in terms of quantitative mass limits, probing the dynamical history and mass evolution of clusters, and the role of dark matter in small bodies. We therefore strongly encourage further studies of the chemical enrichment of these most-massive clusters, in the context of their ICM evolution.

\section{Conclusions}
\label{ConcSect}

We have shown that UV radiation from hot post-AGB stars and cooling white dwarfs can continually ionize the observed ICM of 47 Tucanae, and that these provide the dominant source of ionizing radiation within the cluster. We show that individual stars (e.g. UIT-14) become sufficiently UV-bright to ionize the ICM once they reach $\approx$14\,000 K on their post-AGB tracks, and continue to effectively ionize the ICM for 2--4 Myr while on the white dwarf cooling track. This timescale is sufficiently longer than the 80\,000-year stellar death rate that the cluster is permanently ionized. The inferred UV flux from observations is $\sim$10--60$\times$ greater than that needed to ionize the ICM. However, without imaging near 1000 \AA, it is not currently possible to discern exactly which sources are the major ionizers within the cluster.

We have used stellar population modelling to determine the UV dissociation of ICM in a model of 47 Tucanae, showing that the ionization is a long-term phenomenon that is likely to be present in other clusters, including ones of lower metallicity and lower total mass.

We use a hydrostatic model to show that a thermalized ICM can expand sufficiently that it can overflow the cluster's tidal radius. However, with the ICM cloud truncated at this radius, Jeans escape should be more effective than tidal escape. The ICM responds to this lost material on the dynamical timescale, thus the mass of ICM present should be equal to the mass lost by cluster stars over a dynamical timescale. Our modelled dynamical timescale and clearing timescale for 47 Tuc agree at $\sim$4 Myr, implying $\sim$11 M$_\odot$ of ionized ICM exists within that cluster.

We estimate that this will allow material to be cleared from all extant globular clusters, but may allow ICM to be retained by clusters of $\gtrsim$3 $\times$ 10$^6$ M$_\odot$. The precise value will depend on the mass and density profile of the cluster, and its interaction with the Galactic gravitational field and the surrounding Halo gas. Younger clusters are better able to retain their ICM: clusters are more massive, the mass-loss rate per star is higher and the post-AGB evolution of stars happens faster. This should encourage the growth of multiple stellar populations in the more massive and denser clusters, as is observed. We predict the lowest-mass globular clusters should have few second-generation stars, exhibiting little enrichment. Conversely, a mass of little more that of 47 Tuc may set lower mass limit for `transition objects', which comprise of multiple stellar populations with varying [Fe/H]. Between the bounds of $\sim$10$^6$ and 10$^7$ M$_\odot$ should lie a region in which some clusters were able to retain their ICM but are now longer capable of doing so.


\section*{Acknowledgements}

We would like to express our thanks to Peter van Hoof and Raphael Herschi for their respective expertise in {\sc cloudy} and {\sc mesa}. We also thank Martha Boyer for helpful comments on the manuscript, and the anonymous referee for an exceptionally helpful and erudite review of this manuscript.


\label{lastpage}

\end{document}